\documentclass[twocolumn]{aastex63}
\pdfoutput=1
\usepackage{amsmath}
\usepackage{amsfonts}
\usepackage{amssymb}
\usepackage{url}
\usepackage{xspace}
\usepackage{xcolor}
\usepackage{tikz}
\usepackage{enumitem}

\usepackage[latin1]{inputenc}
\usepackage[T1]{fontenc}
\usepackage{soul}

\usepackage{multirow}
\usepackage{bm}

\newcommand{\todo}{\ifmmode \text{\color{red}\Huge{\(\bullet\)}} \else {\color{red}{\Huge$\bullet$}}\fi}
\newcommand{\tido}{\ifmmode {{\color{red}\bullet}} \else {\color{red}$\bullet$}\fi}

\newcommand{\E        }[1]{\ifmmode 10^{#1} \else $10^{#1}$\fi}
\newcommand{\tE        }[1]{\ifmmode \times10^{#1} \else $\times10^{#1}$\fi}
\newcommand{\til}{\ifmmode \sim \else $\sim$\fi}
\renewcommand{\~} {\ifmmode \sim \else $\sim$\fi}

\newcommand{\pc}	{\ifmmode {\rm pc} \else pc\fi}
\newcommand{\kpc}	{\ifmmode {\rm kpc} \else kpc\fi}
\newcommand{\ld}	{\ifmmode {\rm l.d.} \else l.d.\fi}
\newcommand{\kms}	{\ifmmode {\rm km\,s}^{-1} \else km\,s$^{-1}$\fi}
\newcommand{\cc}	{\ifmmode {\rm cm}^{-3}    \else cm$^{-3}$\fi}
\newcommand{\cmii}	{\ifmmode {\rm cm}^{-2}    \else cm$^{-2}$\fi}
\newcommand{\ergs}	{\ifmmode {\rm erg\,s}^{-1} \else erg s$^{-1}$\fi}
\newcommand{\ergcms}	{\ifmmode {\rm erg\,cm}^{-2}\,{\rm s}^{-1} \else erg\,cm$^{-2}$\,s$^{-1}$\fi}
\newcommand{\ergcmsA}	{\ifmmode {\rm erg\,cm}^{-2}\,{\rm s}^{-1}\,{\rm\AA}^{-1}
\else erg\,cm$^{-2}$\,s$^{-1}$\,\AA$^{-1}$\fi}
\newcommand{  \ergcmsHz  }{\ifmmode{\rm erg\,cm}^{-2}\,{\rm s}^{-1}\,{\rm Hz}^{-1}
                       \else ergs\,cm$^{-2}$\,s$^{-1}$\,Hz$^{-1}$\fi}
\newcommand{\kev}	{\ifmmode {\rm keV} \else keV\fi}

\newcommand{\mic}	{\ifmmode {\rm \mu m} \else $\mu$m\fi}
\newcommand{\vFWHM}	{\ifmmode v_{\mbox{\tiny FWHM}} \else $v_{\mbox{\tiny FWHM}}$\fi}
\newcommand{\vBLR}	{\ifmmode v_{\mbox{\tiny BLR}} \else $v_{\mbox{\tiny BLR}}$\fi}
\newcommand{\sigBLR}	{\ifmmode \sigma_{\mbox{\tiny BLR}} \else $\sigma_{\mbox{\tiny BLR}}$\fi}
\newcommand{\vNLR}	{\ifmmode v_{\mbox{\tiny NLR}} \else $v_{\mbox{\tiny NLR}}$\fi}
\newcommand{\tauBLR}	{\ifmmode \tau_{\mbox{\tiny BLR}} \else $\tau_{\mbox{\tiny BLR}}$\fi}

\newcommand{\Hubble}	{\ifmmode {\rm km\,s}^{-1}\,{\rm Mpc}^{-1} \else km\,s$^{-1}$\,Mpc$^{-1}$\fi}
\newcommand{\NDunit}	{\ifmmode {\rm Mpc}^{-3} \else Mpc$^{-3}$\fi}
\newcommand{\LFunit}	{\ifmmode {\rm Mpc}^{-3}\,{\rm mag}^{-1} \else Mpc$^{-3}$\,mag$^{-1}$\fi}
\newcommand{\MFunit}	{\ifmmode {\rm Mpc}^{-3}\,{\rm dex}^{-1} \else Mpc$^{-3}$\,dex$^{-1}$\fi}

\newcommand{\Msun}{\ifmmode M_{\odot} \else $M_{\odot}$\fi}
\newcommand{\Lsun}{\ifmmode L_{\odot} \else $L_{\odot}$\fi}
\newcommand{\Zsun}{\ifmmode Z_{\odot} \else $Z_{\odot}$\fi}
\newcommand{\mpyr}{\ifmmode \Msun\,{\rm yr}^{-1} \else $\Msun\,{\rm yr}^{-1}$\fi}

\newcommand{\Msol}{\Msun}

\newcommand{\qnote}{\ifmmode q_{0} \else $q_{0}$\fi}
\newcommand{\Hnote}{\ifmmode H_{0} \else $H_{0}$\fi}
\newcommand{\hnote}{\ifmmode h_{0} \else $h_{0}$\fi}
\newcommand{\anote}{\ifmmode a_{0} \else $a_{0}$\fi}
\newcommand{\tnote}{\ifmmode t_{0} \else $t_{0}$\fi}

\def\gsim{\;\rlap{\lower 2.5pt \hbox{$\sim$}}\raise 1.5pt\hbox{$>$}\;}
\def\lsim{\;\rlap{\lower 2.5pt \hbox{$\sim$}}\raise 1.5pt\hbox{$<$}\;}

\newcommand{  \Halpha   }{\ifmmode {\rm H}\alpha \else H$\alpha$\fi}

\newcommand{  \ha       }{\Halpha}
\newcommand{  \Hbeta    }{\ifmmode {\rm H}\beta \else H$\beta$\fi}

\newcommand{  \hb       }{\Hbeta}
\newcommand{  \Hgamma   }{\ifmmode {\rm H}\gamma \else H$\gamma$\fi}
\newcommand{  \Hdelta   }{\ifmmode {\rm H}\delta \else H$\delta$\fi}
\newcommand{  \Lya      }{\ifmmode {\rm Ly}\alpha \else Ly$\alpha$\fi}
\newcommand{  \Lyb      }{\ifmmode {\rm Ly}\beta \else Ly$\beta$\fi}
\newcommand{  \Pa       }{\ifmmode {\rm P}\alpha \else P$\alpha$\fi}
\newcommand{  \Pb       }{\ifmmode {\rm P}\beta \else P$\beta$\fi}
\newcommand{  \Bra      }{\ifmmode {\rm Br}\alpha \else Br$\alpha$\fi}
\newcommand{  \Brg      }{\ifmmode {\rm Br}\gamma \else Br$\gamma$\fi}
\newcommand{  \hii      }{\ifmmode {\rm H}\,\textsc{ii} \else H\,\textsc{ii}\fi}
\newcommand{  \hei      }{\ifmmode {\rm He}\,\textsc{i} \else He\,\textsc{i}\fi}
\newcommand{  \heii     }{\ifmmode {\rm He}\,\textsc{ii} \else He\,\textsc{ii}\fi}
\newcommand{  \HeIIuv   }{\ifmmode {\rm He}\,\textsc{ii}\,\lambda1640 \else He\,\textsc{ii}\,$\lambda1640$\fi}
\newcommand{  \HeIIop   }{\ifmmode {\rm He}\,\textsc{ii}\,\lambda4686 \else He\,\textsc{ii}\,$\lambda4686$\fi}
\newcommand{  \CII	}{\ifmmode \left[{\rm C}\,\textsc{ii}\right]\,\lambda157.74\,\mu{\rm m} \else [C\,{\sc ii}]\ $\lambda157.74\,\mu{\rm m}$\fi}
\newcommand{  \cii	}{\ifmmode \left[{\rm C}\,\textsc{ii}\right] \else [C\,{\sc ii}]\fi}

\newcommand{  \ciii     }{\ifmmode {\rm C}\,\textsc{iii}\right] \else C\,\textsc{iii}]\fi}
\newcommand{  \CIII     }{\ifmmode {\rm C}\,\textsc{iii}\right]\,\lambda1909 \else C\,\textsc{iii}]\,$\lambda1909$\fi}
\newcommand{  \civ      }{\ifmmode {\rm C}\,\textsc{iv}  \else C\,\textsc{iv}\fi}
\newcommand{  \CIV      }{\ifmmode {\rm C}\,\textsc{iv}\,\lambda1549 \else C\,\textsc{iv}\,$\lambda1549$\fi}
\newcommand{  \NIIopt   }{\ifmmode \left[{\rm N}\,\textsc{ii}\right]\,\lambda6584 \else [N\,\textsc{ii}]\,$\lambda6584$\fi}
\newcommand{  \nii      }{\ifmmode \left[{\rm N}\,\textsc{ii}\right]  \else [N\,\textsc{ii}]\fi}
\newcommand{  \niii     }{\ifmmode {\rm N}\,\textsc{iii} \else N\,\textsc{iii}\fi}
\newcommand{  \NIII     }{\ifmmode {\rm N}\,\textsc{iii}\,\lambda4640 \else N\,\textsc{iii}\,$\lambda4640$\fi}
\newcommand{  \niv      }{\ifmmode {\rm N}\,\textsc{iv}  \else N\,\textsc{iv}\fi}
\newcommand{  \NIVuv    }{\ifmmode {\rm N}\,\textsc{iv}\,\lambda1486 \else N\,\textsc{iv}\,$\lambda1486$\fi}
\newcommand{  \nv       }{\ifmmode {\rm N}\,\textsc{v}   \else N\,\textsc{v}\fi}
\newcommand{\oi}{\ifmmode \left[{\rm O}\,\textsc{i}\right] \else [O\,{\sc i}]\fi}
\newcommand{\OI}{\ifmmode \left[{\rm O}\,\textsc{i}\right]\,\lambda6300 \else [O\,{\sc i}]$\,\lambda6300$\fi}
\newcommand{\oii}{\ifmmode \left[{\rm O}\,\textsc{ii}\right] \else [O\,{\sc ii}]\fi}
\newcommand{\OII}{\ifmmode \left[{\rm O}\,\textsc{ii}\right]\,\lambda3727 \else [O\,{\sc ii}]\,$\lambda3727$\fi}
\newcommand{\oiii}{\ifmmode \left[{\rm O}\,\textsc{iii}\right] \else [O\,{\sc iii}]\fi}
\newcommand{\OIII}{\ifmmode \left[{\rm O}\,\textsc{iii}\right]\,\lambda5007 \else [O\,{\sc iii}]\,$\lambda5007$\fi}
\newcommand{  \OIIIbf   }{\ifmmode {\rm O}\,\textsc{iii}\,\lambda3133 \else O\,\textsc{iii}\,$\lambda3133$\fi}
\newcommand{  \OIIIuv   }{\ifmmode {\rm O}\,\textsc{iii}\,\lambda1663 \else O\,\textsc{iii}\,$\lambda1663$\fi}
\newcommand{  \oiv      }{\ifmmode {\rm O}\,\textsc{iv}  \else O\,\textsc{iv}\fi}
\newcommand{  \OIVuv    }{\ifmmode {\rm O}\,\textsc{iv}\,\lambda1402  \else O\,\textsc{iv}\,$\lambda1402$\fi}
\newcommand{  \OIVIR    }{\ifmmode {\rm O}\,\textsc{iv}\,25.9\,\mu {\rm m} \else O\,\textsc{iv}\,$25.9\,\mu$m\fi}
\newcommand{  \ovi      }{\ifmmode {\rm O}\,\textsc{vi}   \else O\,\textsc{vi}\fi}
\newcommand{  \Ovi      }{\ifmmode {\rm O}\,\textsc{vi}\,\lambda1035 \else O\,\textsc{vi}\,$\lambda1035$\fi}
\newcommand{  \nei      }{\ifmmode {\rm Ne}\,\textsc{i}   \else Ne\,\textsc{i}\fi}
\newcommand{  \neii     }{\ifmmode {\rm Ne}\,\textsc{ii}  \else Ne\,\textsc{ii}\fi}
\newcommand{  \NeiiIR   }{\ifmmode {\rm Ne}\,\textsc{ii}\,12.8\,\mu {\rm m} \else Ne\,\textsc{ii}\,$12.8\,\mu$m\fi}
\newcommand{  \neiii    }{\ifmmode {\rm Ne}\,\textsc{iii} \else Ne\,\textsc{iii}\fi}
\newcommand{  \neiv     }{\ifmmode {\rm Ne}\,\textsc{iv}  \else Ne\,\textsc{iv}\fi}
\newcommand{  \nev      }{\ifmmode \left[{\rm Ne}\,\textsc{v}\right]   \else [Ne\,\textsc{v}]\fi}
\newcommand{  \NevIR    }{\ifmmode \left[{\rm Ne}\,\textsc{v}\right]\,\lambda24.3\,\mu {\rm m} \else Ne\,\textsc{v}\,$\lambda24.3\,\mu$m\fi}
\newcommand{  \nevi     }{\ifmmode {\rm Ne}\,\textsc{vi}  \else Ne\,\textsc{vi}\fi}
\newcommand{  \mgi      }{\ifmmode {\rm Mg}\,\textsc{i} \else Mg\,\textsc{i}\fi}
\newcommand{  \mgii     }{\ifmmode {\rm Mg}\,\textsc{ii} \else Mg\,\textsc{ii}\fi}
\newcommand{  \MgII     }{\ifmmode {\rm Mg}\,\textsc{ii}\,\lambda2798 \else Mg\,\textsc{ii}\,$\lambda2798$\fi}
\newcommand{  \sii      }{\ifmmode {\rm S}\,\textsc{ii} \else S\,\textsc{ii}\fi}
\newcommand{  \siii     }{\ifmmode {\rm S}\,\textsc{iii} \else S\,\textsc{iii}\fi}
\newcommand{  \siv      }{\ifmmode {\rm S}\,\textsc{iv} \else S\,\textsc{iv}\fi}
\newcommand{  \sili     }{\ifmmode {\rm Si}\,\textsc{i}   \else Si\,\textsc{i}\fi}
\newcommand{  \silii    }{\ifmmode {\rm Si}\,\textsc{ii}  \else Si\,\textsc{ii}\fi}
\newcommand{  \Siliv    }{\ifmmode {\rm Si}\,\textsc{iv}  \else Si\,\textsc{iv}\fi}
\newcommand{  \SilIVuv  }{\ifmmode {\rm Si}\,\textsc{iv}\,\lambda1400  \else Si\,\textsc{iv}\,$\lambda1400$\fi}
\newcommand{  \AlIII   }{\ifmmode {\rm Al}\,\textsc{iii}\,\lambda1857 \else Al\,\textsc{iii}\,$\lambda1857$\fi}
\newcommand{  \Aliii   }{\ifmmode {\rm Al}\,\textsc{iii} \else Al\,\textsc{iii}\fi}
\newcommand{  \caii     }{\ifmmode {\rm Ca}\,\textsc{ii} \else Ca\,\textsc{ii}\fi}
\newcommand{  \feii     }{\ifmmode {\rm Fe}\,\textsc{ii} \else Fe\,\textsc{ii}\fi}
\newcommand{  \feiii    }{\ifmmode {\rm Fe}\,\textsc{iii} \else Fe\,\textsc{iii}\fi}
\newcommand{  \Kalpha   }{\ifmmode {\rm K}\alpha \else K$\alpha$\fi}

\newcommand{ \Lhb   }{\ifmmode L_{\hb} \else $L_{\hb}$\fi}
\newcommand{ \Lha   }{\ifmmode L_{\ha} \else $L_{\ha}$\fi}
\newcommand{ \fwhb  }{\ifmmode {\rm FWHM}\left(\hb\right) \else FWHM(\hb)\fi}
\newcommand{\sighb  }{\ifmmode \sigma\left(\hb\right) \else $\sigma\left(\hb\right)$\fi}
\newcommand{ \ewhb  }{\ifmmode {\rm EW}\left(\hb\right) \else EW(\hb)\fi}
\newcommand{ \fwha  }{\ifmmode {\rm FWHM}\left(\ha\right) \else FWHM(\ha)\fi}
\newcommand{ \ewha  }{\ifmmode {\rm EW}\left(\ha\right) \else EW(\ha)\fi}
\newcommand{ \Lmg   }{\ifmmode L\left(\mgii\right) \else $L\left(\mgii\right)$\fi}
\newcommand{ \fwmg  }{\ifmmode {\rm FWHM}\left(\mgii\right) \else FWHM(\mgii)\fi}
\newcommand{ \Lciv  }{\ifmmode L\left(\civ\right) \else $L\left(\civ\right)$\fi}
\newcommand{ \fwciv }{\ifmmode {\rm FWHM}\left(\civ\right) \else FWHM(\civ)\fi}
\newcommand{ \fwhm  }{\ifmmode {\rm FWHM} \else FWHM\fi} 
\newcommand{ \voff  }{\ifmmode v_{\rm off} \else $v_{\rm off}$\fi} 
\newcommand{ \vmax  }{\ifmmode v_{\rm max} \else $v_{\rm max}$\fi} 

\newcommand{ \mumg  }{\ifmmode \mu\left(\mgii\right) \else $\mu\left(\mgii\right)$\fi}
\newcommand{ \fmg   }{\ifmmode f\left(\mgii\right) \else $f\left(\mgii\right)$\fi}
\newcommand{ \muciv }{\ifmmode \mu\left(\civ\right) \else $\mu\left(\civ\right)$\fi}
\newcommand{ \fciv  }{\ifmmode f\left(\civ\right) \else $f\left(\civ\right)$\fi}
\newcommand{  \auvo     }{\ifmmode \alpha_{\nu,{\rm UVO}} \else $\alpha_{\nu,{\rm UVO}}$\fi}
\newcommand{  \Ledd     }{\ifmmode L_{\rm Edd} \else $L_{\rm Edd}$\fi}
\newcommand{  \lamLlam  }{\ifmmode \lambda L_{\lambda} \else $\lambda L_{\lambda}$\fi}
\newcommand{  \lLl      }{\ifmmode \lambda L_{\lambda} \else $\lambda L_{\lambda}$\fi}
\newcommand{  \nuLnu    }{\ifmmode \nu L_{\nu} \else $\nu L_{\nu}$\fi}
\newcommand{  \nLn      }{\ifmmode \nu L_{\nu} \else $\nu L_{\nu}$\fi}
\newcommand{  \Luv      }{\ifmmode L_{1450} \else $L_{1450}$\fi}
\newcommand{  \Lop      }{\ifmmode L_{5100} \else $L_{5100}$\fi}
\newcommand{  \lLop     }{\ifmmode \log\left(\Lop/\ergs\right) \else $\log\left(\Lop/\ergs\right)$\fi}
\newcommand{  \Lthree   }{\ifmmode L_{3000} \else $L_{3000}$\fi}
\newcommand{  \lLthree  }{\ifmmode \log\left(\Lthree/\ergs\right) \else $\log\left(\Lthree/\ergs\right)$\fi}
\newcommand{  \Lsix      }{\ifmmode L_{6200} \else $L_{6200}$\fi}
\newcommand{  \lLisx     }{\ifmmode \log\left(\Lop/\ergs\right) \else $\log\left(\Lop/\ergs\right)$\fi}
\newcommand{  \Lhard    }{\ifmmode L_{\rm 2-10} \else $L_{\rm 2-10}$\fi}
\newcommand{  \Fhard    }{\ifmmode F_{\rm 2-10} \else $F_{\rm 2-10}$\fi}
\newcommand{  \Lsoft    }{\ifmmode L_{\rm 0.5-2} \else $L_{\rm 0.5-2}$\fi}
\newcommand{  \Luhard    }{\ifmmode L_{\rm 14-150} \else $L_{\rm 14-150}$\fi}
\newcommand{  \Fuhard    }{\ifmmode F_{\rm 14-150} \else $F_{\rm 14-150}$\fi}
\newcommand{  \Lbat    }{\ifmmode L_{\rm 14-195} \else $L_{\rm 14-195}$\fi}
\newcommand{  \Fbat    }{\ifmmode F_{\rm 14-195} \else $F_{\rm 14-195}$\fi}
\newcommand{  \Lxray    }{\ifmmode L_{\rm X} \else $L_{\rm X}$\fi}
\newcommand{  \Fxray    }{\ifmmode F_{\rm X} \else $F_{\rm X}$\fi}
\newcommand{  \Lx    }{\Lxray}

\newcommand{\Fthree}{\ifmmode F_{3000} \else $F_{3000}$\fi}
\newcommand{\fuv}{\ifmmode f_{\lambda}\left(1450{\rm \AA}\right) \else $f_{\lambda}\left(1450 {\rm \AA}\right)$\fi}
\newcommand{\fthree}{\ifmmode f_{\lambda}\left(3000{\rm \AA}\right) \else $f_{\lambda}\left(3000{\rm \AA}\right)$\fi}
\newcommand{\fH}{\ifmmode f_{\lambda}\left(1.65\micron\right) \else
$f_{\lambda}\left(1.65\micron\right)$\fi}

\newcommand{\fbol}{\ifmmode f_{\rm bol} \else $f_{\rm bol}$\fi}
\newcommand{\fbolwv}{\ifmmode f_{\rm bol}\left(\lambda\right) \else $f_{\rm bol}\left(\lambda\right)$\fi}
\newcommand{\fbolopt}{\ifmmode f_{\rm bol}\left(5100{\rm \AA}\right) \else $f_{\rm bol}\left(5100{\rm \AA}\right)$\fi}
\newcommand{\fbolthree}{\ifmmode f_{\rm bol}\left(3000{\rm \AA}\right) \else $f_{\rm bol}\left(3000{\rm \AA}\right)$\fi}
\newcommand{\fboluv}{\ifmmode f_{\rm bol}\left(1450{\rm \AA}\right) \else $f_{\rm bol}\left(1450{\rm \AA}\right)$\fi}

\newcommand{\fbolbat}{\ifmmode f_{\rm bol}\left(14-150\,\kev\right) \else $f_{\rm bol}\left(14-150\,\kev\right)$\fi}
\newcommand{\fbolhard}{\ifmmode f_{\rm bol}\left(2-10\,\kev\right) \else $f_{\rm bol}\left(2-10\,\kev\right)$\fi}

\newcommand{\fobs}{\ifmmode f_{\rm obs} \else $f_{\rm obs}$\fi}

\newcommand{  \mbh      }{\ifmmode M_{\rm BH} \else $M_{\rm BH}$\fi}
\newcommand{  \lmbh     }{\ifmmode \log\left(\mbh/\Msun\right) \else $\log\left(\mbh/\Msun\right)$\fi} 
\newcommand{  \lledd    }{\ifmmode L/L_{\rm Edd} \else $L/L_{\rm Edd}$\fi}
\newcommand{  \mmedd    }{\ifmmode \dot{m}/\dot{m}_{\rm \,Edd} \else $\dot{m}/\dot{m}_{\rm \,Edd}$\fi}
\newcommand{  \Lbol     }{\ifmmode L_{\rm bol} \else $L_{\rm bol}$\fi}
\newcommand{  \lbol     }{\ifmmode L_{\rm bol} \else $L_{\rm bol}$\fi}
\newcommand{  \lLbol    }{\ifmmode \log\left(\Lbol/\ergs\right) \else $\log\left(\Lbol/\ergs\right)$\fi} 
\newcommand{  \Lagn     }{\ifmmode L_{\rm AGN} \else $L_{\rm AGN}$\fi}
\newcommand{  \lagn     }{\ifmmode L_{\rm AGN} \else $L_{\rm AGN}$\fi}

\newcommand{  \tgrow     }{\ifmmode t_{\rm growth} \else $t_{\rm growth}$\fi}
\newcommand{  \tAD     }{\ifmmode t_{\rm acc} \else $t_{\rm acc}$\fi}
\newcommand{  \tacc    }{\ifmmode t_{\rm acc} \else $t_{\rm acc}$\fi}
\newcommand{  \tUni      }{\ifmmode t_{\rm Universe} \else $t_{\rm Universe}$\fi}

\newcommand{  \Mdotin	}{\ifmmode \dot{M}_{\rm infall} \else $\dot{M}_{\rm infall}$\fi}
\newcommand{  \Mdotbh	}{\ifmmode \dot{M}_{\rm BH} \else $\dot{M}_{\rm BH}$\fi}
\newcommand{  \Mdotad	}{\ifmmode \dot{M}_{\rm AD} \else $\dot{M}_{\rm AD}$\fi}
\newcommand{  \Mdotacc	}{\ifmmode \dot{M}_{\rm acc} \else $\dot{M}_{\rm acc}$\fi}
\newcommand{  \Mdotthin	}{\ifmmode \dot{M}_{\rm thin} \else $\dot{M}_{\rm thin}$\fi}
\newcommand{  \Mdotdisk	}{\ifmmode \dot{M}_{\rm disk} \else $\dot{M}_{\rm disk}$\fi}

\newcommand{  \Mindot	}{\ifmmode \dot{M}_{\rm infall} \else $\dot{M}_{\rm infall}$\fi}
\newcommand{  \Mbhdot	}{\ifmmode \dot{M}_{\rm BH} \else $\dot{M}_{\rm BH}$\fi}
\newcommand{  \Maddot	}{\ifmmode \dot{M}_{\rm AD} \else $\dot{M}_{\rm AD}$\fi}
\newcommand{  \Maccdot	}{\ifmmode \dot{M}_{\rm acc} \else $\dot{M}_{\rm acc}$\fi}
\newcommand{  \Mthdot	}{\ifmmode \dot{M}_{\rm thin} \else $\dot{M}_{\rm thin}$\fi}
\newcommand{  \Mdsdot	}{\ifmmode \dot{M}_{\rm disk} \else $\dot{M}_{\rm disk}$\fi}

\newcommand{  \as	}{\ifmmode a_{\rm *} \else $a_{\rm *}$\fi}
\newcommand{  \avec	}{\ifmmode \vec{a}_{\rm *} \else $\vec{a}_{\rm *}$\fi}
\newcommand{  \re	}{\ifmmode \eta      	 \else $\eta$\fi}
\newcommand{  \RISCO	}{\ifmmode R_{\rm ISCO}  \else $R_{\rm ISCO}$\fi}

\newcommand{  \mseed    }{\ifmmode M_{\rm seed} \else $M_{\rm seed}$\fi}
\newcommand{  \mbul     }{\ifmmode M_{\rm bulge} \else $M_{\rm bulge}$\fi} 
\newcommand{  \mstar    }{\ifmmode M_{*} \else $M_{*}$\fi} 
\newcommand{  \mgal     }{\ifmmode M_{*} \else $M_{*}$\fi} 
\newcommand{  \mhost    }{\ifmmode M_{\rm host} \else $M_{\rm host}$\fi}
\newcommand{  \mmsmall  }{\ifmmode M_{\rm BH}/M_{*} \else $M_{\rm BH}/M_{*}$\fi}
\newcommand{  \mmlarge  }{\ifmmode M_{*}/M_{\rm BH} \else $M_{*}/M_{\rm BH}$\fi}

\newcommand{  \mmdotlarge}{\ifmmode \dot{M}_*/\Mbhdot \else $\dot{M}_*/\Mbhdot$\fi}
\newcommand{  \mmdotsmall}{\ifmmode \Mbhdot/\dot{M}_* \else $\Mbhdot/\dot{M}_*$\fi}

\newcommand{  \mmwp     }{\ifmmode \left(M_{*}/M_{\rm BH}\right) \else $\left(M_{*}/M_{\rm BH}\right)$\fi}
\newcommand{  \ml       }{\ifmmode M_{*}/L_{*} \else $M_{*}/L_{*}$\fi}
\newcommand{  \mlwp     }{\ifmmode \left(M_{*}/L\right) \else $\left(M_{*}/L\right)$\fi}
\newcommand{  \mlk      }{\ifmmode \left(M_{*}/L_{K}\right) \else $\left(M_{*}/L_{K}\right)$\fi}
\newcommand{  \sigs     }{\ifmmode \sigma_{*} \else $\sigma_{*}$\fi}
\newcommand{  \Reff     }{\ifmmode R_{\rm e} \else $R_{\rm e}$\fi}
\newcommand{  \Rvir     }{\ifmmode R_{\rm vir} \else $R_{\rm vir}$\fi}
\newcommand{  \Rtwo     }{\ifmmode R_{200} \else $R_{200}$\fi}
\newcommand{  \Rfive    }{\ifmmode R_{500} \else $R_{500}$\fi}
\newcommand{  \Rgrp     }{\ifmmode R_{\rm grp} \else $R_{\rm grp}$\fi}
\newcommand{  \nser     }{\ifmmode n_{\rm s} \else $n_{\rm s}$\fi}
\newcommand{  \LSF      }{\ifmmode L_{\rm SF}  \else $L_{\rm SF}$\fi}
\newcommand{  \LFIR     }{\ifmmode L_{\rm FIR} \else $L_{\rm FIR}$\fi}
\newcommand{  \Lfir     }{\ifmmode L_{\rm FIR} \else $L_{\rm FIR}$\fi}
\newcommand{  \LTIR     }{\ifmmode L_{\rm TIR} \else $L_{\rm TIR}$\fi}
\newcommand{  \Ltir     }{\ifmmode L_{\rm TIR} \else $L_{\rm TIR}$\fi}

\newcommand{  \mdyn     }{\ifmmode M_{\rm dyn} \else $M_{\rm dyn}$\fi} 
\newcommand{  \mgas     }{\ifmmode M_{\rm gas} \else $M_{\rm gas}$\fi} 
\newcommand{  \mh       }{\ifmmode M_{\rm h} \else $M_{\rm h}$\fi}
\newcommand{  \mhalo    }{\ifmmode M_{\rm halo} \else $M_{\rm halo}$\fi}
\newcommand{  \sfr      }{\ifmmode {\rm SFR} \else SFR\fi}

\newcommand{ \Lcii     }{\ifmmode L_{\cii} \else $L_{\cii}$\fi}
\newcommand{ \fwcii  }{\ifmmode {\rm FWHM}\cii \else FWHM\cii\fi}

\newcommand{\bj}{\ifmmode b_{\rm J} \else $b_{\rm J}$\fi}

\newcommand{\iab}{\ifmmode i_{\rm AB} \else $i_{\rm AB}$\fi}

\newcommand{\jab}{\ifmmode J_{\rm AB} \else $J_{\rm AB}$\fi}
\newcommand{\hab}{\ifmmode H_{\rm AB} \else $H_{\rm AB}$\fi}
\newcommand{\kab}{\ifmmode K_{\rm AB} \else $K_{\rm AB}$\fi}

\newcommand{\jveg}{\ifmmode J_{\rm Vega} \else $J_{\rm Vega}$\fi}
\newcommand{\hveg}{\ifmmode H_{\rm Vega} \else $H_{\rm Vega}$\fi}
\newcommand{\kveg}{\ifmmode K_{\rm Vega} \else $K_{\rm Vega}$\fi}

\def\arcsec{\hbox{$^{\prime\prime}$}}

\newcommand{  \Chisq    }{\ifmmode \chi^{2} \else $\chi^{2}$}
\newcommand{  \nelec    }{\ifmmode n_{e} \else $n_{e}$\fi}     % electron density
\newcommand{  \nh       }{\ifmmode n_{\rm H} \else $n_{\rm H}$\fi}     % hydrogen density
\newcommand{  \Ncol     }{\ifmmode N_{\rm col} \else $N_{\rm col}$\fi} % column density
\newcommand{  \NH       }{\ifmmode N_{\rm H} \else $N_{\rm H}$\fi}     % column density

\def\deg{\hbox{$^\circ$}}
\def\sun{\hbox{$\odot$}}

\def\arcsec{\hbox{$^{\prime\prime}$}}
\def\ion#1#2{#1$\;${\small\rm\@Roman{#2}}\relax}

\newcommand{\NeV}{\ifmmode \left[{\rm Ne}\,\textsc{v}\right]\,\lambda3427 \else [Ne\,\textsc{v}]\,$\lambda3427$\fi}
\newcommand{\lamEdd}{\ifmmode \lambda_{\rm Edd} \else $\lambda_{\rm Edd}$\fi}
\newcommand{\lognh}{\ifmmode \log(\NH/\cmii) \else $\log(\NH/\cmii)$\fi}
\newcommand{\fnev }{\ifmmode F_{\nev} \else $F_{\nev}$\fi}
\newcommand{\Lnev }{\ifmmode L_{\nev} \else $L_{\nev}$\fi}
\newcommand{\ewnev}{\ifmmode {\rm EW}_{\nev} \else EW$_{\nev}$\fi}
\newcommand{\Loiii}{\ifmmode L_{\oiii} \else $L_{\oiii}$\fi}
\newcommand{\fdet}{\ifmmode f_{\rm det} \else $f_{\rm det}$\fi}
\newcommand{\ffx}{\ifmmode \fnev/\Fuhard \else $\fnev/\Fuhard$\fi}

\newcommand{\Nnev}{341}
\newcommand{\Ndet}{146}

\newcommand{\Nnh}{234}
\newcommand{\Nmbh}{276}

\newcommand{\chm}{\checkmark}

\received{December 20, 2024}
\revised{June 11, 2025}
\accepted{June 24, 2025}
\submitjournal{ApJ}

\shorttitle{BASS XLVIII: the \NeV\ emission line}
\shortauthors{Reiss et al.}
\graphicspath{{./}{figures/}}

\begin{document}

%%%%%%%%%%%%%%%%%%%%%%%%%%%%%%%%%%%%%%%%%%%%%%%%%%%%%%%%%%%%%%%%%%%%%%%%%%%%%%%%
\title{BASS XLVIII: \NeV\ Emission in Powerful Nearby Active Galactic Nuclei}
\correspondingauthor{Benny Trakhtenbrot}
\email{bennyt@tauex.tau.ac.il}
%%%%%%%%%%%%%%%%%%%%%%%%%%%%%%%%%%%%%%%%%%%%%%%%%%%%%%%%%%%%%%%%%%%%%%%%%%%%%%%%
\author{Tomer Reiss}
\affiliation{School of Physics and Astronomy, Tel Aviv University, Tel Aviv 69978, Israel}

\author[0000-0002-3683-7297]{Benny Trakhtenbrot}
\affiliation{School of Physics and Astronomy, Tel Aviv University, Tel Aviv 69978, Israel}
\affiliation{Max-Planck-Institut f{\"u}r extraterrestrische Physik, Gie\ss{}enbachstra\ss{}e 1, 85748 Garching, Germany}
\affiliation{Excellence Cluster ORIGINS, Boltzmannsstra\ss{}e 2, 85748, Garching, Germany}

\author[0000-0001-5231-2645]{Claudio Ricci}
\affiliation{Instituto de Estudios Astrof\'isicos, Facultad de Ingenier\'ia y Ciencias, Universidad Diego Portales, Av. Ej\'ercito Libertador 441, Santiago, Chile} 
\affiliation{Kavli Institute for Astronomy and Astrophysics, Peking University, Beijing 100871, China}

\author[0000-0002-8686-8737]{Franz E. Bauer}
\affiliation{Instituto de Alta Investigaci{\'{o}}n, Universidad de Tarapac{\'{a}}, Casilla 7D, Arica, Chile}
% \affiliation{Instituto de Astrof\'isica, Pontificia Universidad Cat\'olica de Chile, Casilla 306, Santiago 22, Chile}
% \affiliation{Millennium Institute of Astrophysics (MAS), Nuncio Monse$\tilde{\rm n}$or S\'otero Sanz 100, Providencia, Santiago, Chile}
% \affiliation{Space Science Institute, 4750 Walnut Street, Suite 205, Boulder, CO 80301, USA}

\author[0000-0002-7998-9581]{Michael J. Koss}
\affiliation{Eureka Scientific, 2452 Delmer Street, Suite 100, Oakland, CA 94602-3017, USA} 

\author[0000-0002-4377-903X]{Kohei Ichikawa}
\affiliation{Astronomical Institute, Tohoku University, Aramaki, Aoba-ku, Sendai, Miyagi 980-8578, Japan}
\affiliation{Frontier Research Institute for Interdisciplinary Sciences, Tohoku University, Sendai 980-8578, Japan}
\affiliation{Global Center for Science and Engineering, Faculty of
Science and Engineering, Waseda University, 3-4-1, Okubo, Shinjuku,
Tokyo 169-8555, Japan}

\author[0000-0002-2603-2639]{Darshan Kakkad}
\affiliation{Centre for Astrophysics Research, University of Hertfordshire, Hatfield, AL10 9AB, UK}

\author[0000-0002-7962-5446]{Richard Mushotzky}
\affiliation{Department of Astronomy, University of Maryland, College Park, MD 20742, USA} 
\affiliation{Joint Space-Science Institute, University of Maryland, College Park, MD 20742, USA}

\author[0000-0002-5037-951X]{Kyuseok Oh}
\affiliation{Korea Astronomy \& Space Science institute, 776, Daedeokdae-ro, Yuseong-gu, Daejeon 34055, Republic of Korea}

\author[0000-0003-2196-3298]{Alessandro Peca}
\affiliation{Eureka Scientific, 2452 Delmer Street, Suite 100, Oakland, CA 94602-3017, USA}
\affiliation{Department of Physics, Yale University, P.O. Box 208120, New Haven, CT 06520, USA}

\author[0000-0001-5481-8607]{Rudolf B\"ar}
\affiliation{Institute for Particle Physics and Astrophysics, ETH Z\"urich, Wolfgang-Pauli-Strasse 27, CH-8093 Z\"urich, Switzerland}

\author[0000-0002-8604-1158]{Yaherlyn Diaz}
\affiliation{Instituto de Estudios Astrof\'isicos, Facultad de Ingenier\'ia y Ciencias, Universidad Diego Portales, Av. Ej\'ercito Libertador 441, Santiago, Chile}

\author[0000-0002-4226-8959]{Fiona Harrison}
\affiliation{Cahill Center for Astronomy and Astrophysics, California Institute of Technology, Pasadena, CA 91125, USA}

\author[0000-0003-2284-8603]{Meredith C. Powell}
\affiliation{Leibniz-Institut f\"ur Astrophysik Potsdam (AIP), An der Sternwarte 16, 14482 Potsdam, Germany}
\affiliation{Kavli Institute for Particle Astrophysics and Cosmology, Stanford University, 452 Lomita Mall, Stanford, CA 94305, USA}
 
\author[0000-0002-3140-4070]{Eleonora Sani}
\affiliation{European Southern Observatory, Alonso de C{\'o}rdova 3107, Casilla 19, Santiago 19001, Chile}

\author[0000-0003-2686-9241]{Daniel Stern}
\affiliation{Jet Propulsion Laboratory, California Institute of Technology, 4800 Oak Grove Drive, MS 169-224, Pasadena, CA 91109, USA}

\author[0000-0002-0745-9792]{C. Megan Urry}
\affiliation{Department of Physics, Yale University, P.O.\ Box 208120, New Haven, CT 06520, USA}
\affiliation{Yale Center for Astronomy and Astrophysics, 52 Hillhouse Avenue, New Haven, CT 06511, USA}

%%%%%%%%%%%%%%%%%%%%%%%%%%%%%%%%%%%%%

\begin{abstract}
We investigate the high-ionization, narrow \NeV\ emission line in a sample of over 340 ultra-hard X-ray (14-195 \kev) selected Active Galactic Nuclei (AGN) drawn from the BASS project. 
The analysis includes measurements in individual and stacked spectra, and considers several key AGN properties such as X-ray luminosity, supermassive black hole (SMBH) mass, Eddington ratios, and line-of-sight column density. 
The \NeV\ line is robustly detected in $\approx$43\% (146/341) of the AGN in our sample, with no significant trends between the detection rate and key AGN/SMBH properties. In particular, the detection rate remains high even at the highest levels of obscuration ($>70\%$ for $\log[N_{\rm H}/\cmii]\gtrsim23$). On the other hand, even some of our highest signal-to-noise spectra ($S/N>50$) lack a robust \nev\ detection.
The typical (median) scaling ratios between \nev\ line emission and (ultra-)hard X-ray emission in our sample are $\log\Lnev/L_{14-150\,\kev}\simeq-3.75$ and $\log\Lnev/L_{2-10\,\kev}\simeq-3.36$. The scatter on these scaling ratios,  of ${\lesssim}0.5$\,dex, is comparable to, and indeed smaller than, what is found for other commonly used tracers of AGN radiative outputs (e.g., \OIII).
Otherwise, we find no significant relations between the (relative) strength of \nev\ and the basic AGN/SMBH properties under study, in contrast with simple expectations from models of SMBH accretion flows.
Our results reaffirm the usability of \nev\ as an AGN tracer even in highly obscured systems, including dual AGN and high redshift sources.
\end{abstract}


\section{Introduction} 
\label{sec:intro}

Active galactic nuclei (AGN) are known to emit copious amounts of radiation across the electromagnetic spectrum, making them observable to extremely high redshifts. 
Among their many radiative signatures, some are unique to--or exceptionally dominant---in AGN, which allows complete and pure selection of AGN and the determination of their intrinsic radiative power in large extra-galactic surveys.
Moreover, there is particular interest in those radiative probes that are detectable even in the most obscured sources, and/or that are linked to other key properties of the AGN or of the supermassive black holes (SMBHs) that power them, such as mass or accretion rate. 

In about half of the AGN in the universe the UV-optical emission from the central engine is obscured, making their study particularly challenging. Among the alternative radiative signatures used to identify and understand such sources, two have been used extensively: broad-band, hard X-ray continuum emission, and narrow emission lines from high-ionization species (see review of obscured AGN by \citealt{HickoxAlexander18_Rev}). 
While the former is thought to be directly linked to the central engine, the latter originates from low-density gas on larger scales, photoionized by the hard AGN radiation.

Strong line ratio diagnostics have been used extensively to identify AGN in large spectroscopic surveys \cite[see review by][]{Kewley2019_rev} and to allow significant progress in studies of the populations of AGN and of their host galaxies \cite[e.g.,][]{Kauffmann2003,Brinchmann2004,Schawinski2007,Netzer2009}. %,Salehirad2022}. 
Such diagnostics are based on the fact that the continuum radiation emitted from AGN is typically much harder than that of young stars (in star-forming regions).
The prominent \OIII\ emission line is commonly used not only to identify AGN, but also to infer their bolometric radiative output of AGN (\Lbol), and the underlying SMBH accretion rate, thanks to the link between $L(\oiii)$ and \Lbol\ \citep{Heckman05}. 
However, as this line requires ionizing radiation with $h\nu >35$ eV, it can indeed be contaminated by star formation (SF). More detailed analyses \cite[e.g.,][]{Netzer2009} tried to disentangle the AGN and SF contributions to \OIII\ in an attempt to assess both the AGN and host-galaxy properties of large AGN samples.

The \NeV\ emission line requires ionizing radiation with $h\nu > 97$ eV and is observationally accessible even for low-redshift sources. This makes it more suitable for telling apart AGN and inactive (SF) galaxies, as was clearly demonstrated and utilized in several large galaxy samples. 
For example, the \nev\ line allowed the identification of luminous, narrow-line (obscured) AGN among the huge extragalactic SDSS sample, out to $z\approx0.8$ \cite[``type-II'' quasars; e.g.,][]{Zakamska2003,Yuan2016}. 
The more recent study by \cite{Negus2023} studied \nev-emitting galaxies with spatially-resolved spectroscopy available through the SDSS-IV/MaNGA survey, finding that $\geq90\%$ of such sources would be classified as AGN by other well-established criteria.
Other studies went further, and demonstrated how \nev\ can be used to identify even highly obscured AGN, with line-of-sight column densities of $\log(\NH/\cmii)>23$, out to $z\sim1-2$ \cite[e.g.,][]{Gilli2010}. 
\nev\ was indeed used to study several samples of obscured AGN at intermediate redshifts \cite[e.g.,][]{Vignali2010,Mignoli2013,Lanzuisi2015,Vergani2018}.
Most recently, \cite{Li2024_NeV} showed that \nev\ can be detected in the stacked spectra of mid-IR (MIR) selected AGN, which again probe the obscured population.
Such obscured sources are particularly important for understanding AGN demographics and evolution, as at least $\sim30\%$ of local AGN are thought to be highly obscured \cite[i.e., Compton thick; see, e.g.,][]{Ricci2015_CT}, and this fraction may be yet higher at higher redshifts \cite[e.g.,][]{Merloni2014,Aird2015,Buchner2015,Peca2023}, where gas- and dust-rich major galaxy mergers can further enhance the obscuration of luminous AGN \citep{Ricci2017_mergers,Blecha2018}.

Since \nev\ emission is (almost always) driven by the high-ionization radiation emerging from SMBH accretion flows, one may expect it to be linked to key AGN and SMBH properties. 
Several studies found that \nev\ emission is indeed strongly correlated with the (hard) X-ray emission of AGN.
Specifically, \cite{Gilli2010} found that the typical ratio between \nev\ and observed hard X-ray (2-10 \kev) emission in unobscured (broad-line) AGN is $60-6000$ (typically ${\sim}400$), while for higher-obscuration sources it can drop to lower values ($1-1000$). \cite{Gilli2010} further suggested that the latter, lower ratios can be used to identify highly obscured sources at significant redshifts, as in such sources the observed X-ray fluxes are suppressed relative to the (unattenuated) narrow emission lines, which may originate from larger scales \cite[see, e.g.,][for particularly extended \nev\ emission regions]{Lintott2009,Keel2012}.
\cite{Berney2015} found a statistically significant, but mild, correlation between ultra-hard X-ray (14-195 \kev) emission and \nev\ strength, with a significant scatter of $\sim$0.5\,dex, among a sample of $\sim$50 local AGN, selected through their ultra-hard X-ray emission.
Other studies of such sources investigated the mid-IR lines of \nev\ at 14.3 and/or 24.3 \mic, and generally found extremely strong and tight correlations \cite[scatter of $\lesssim$0.3\,dex; e.g.,][]{Weaver2010,Spinoglio2022}. The fact that these MIR lines are more tightly correlated with the central AGN radiative power, compared with \NeV, may be understood as indicative of the host-galaxy scale (dust) attenuation, which is generally unknown, but should affect the latter much more than the former.

On the other hand, any relations between \nev\ emission and SMBH mass (\mbh), accretion rate (in terms of the Eddington ratio), or other key AGN properties, are yet to be established, although they may be expected based on our basic picture of AGN accretion flows.
Specifically, within the thin accretion disk paradigm, the emergent continuum is expected to shift towards the extreme UV (EUV; ${\approx}10-100$ eV) as the SMBH mass (\mbh) decreases, and/or as the SMBH spin increases.
Moreover, several modifications of the simple thin disk model suggest that the EUV emission could be either significantly enhanced, through additional Comptonization in a so-called ``slim'' disk \cite[e.g.,][]{Done2012,KubotaDone2019}, or be considerably quenched if the inner parts of the accretion flow are advection dominated \cite[e.g.,][]{Ohsuga2005,Pognan2020}.
If any of this happens in real AGN, then emission lines from high-ionization species, such as \nev, could potentially provide a sought-after diagnostic of super-Eddington accretion in SMBHs. This could be extremely important for understanding the first SMBHs to form in the early Universe \cite[e.g.,][]{Madau2014,Volonteri2015}, as probed by the highest-redshift AGN known, at $z\gtrsim6$ \cite{Wang21,Matsuoka22,Bogdan24,Lambrides24}. 

It is important to note that \nev\ emission in extragalactic sources is not exclusive to AGN. Indeed, galaxies with populations of extremely hot (young and/or Wolf-Rayet) stars, or harboring supernova-driven shocks and outflows, and/or accretion onto stellar-mass BHs may produce detectable levels of \nev\ emission. 
This has been demonstrated both by radiative transfer calculations \cite[e.g.,][]{AbelSatyapal2008,Simmonds2021,Cleri2023_diag}, and by samples of galaxies which show \nev\ emission, yet no (or limited) evidence for the kind of AGN activity that would account for the observed \nev\ emission. The latter are typically blue, compact and metal-poor galaxies in the low-redshift universe \cite[e.g.,][]{Izotov2004,Izotov2021} or also higher-redshift systems, perhaps with analogous properties \cite[e.g.,][]{Cleri2023_CLEAR}. 
In addition, several works explicitly examined extended \nev\ emission, which can be ascribed to shocks \cite[e.g.,][]{Leung2021,Negus2023}.

To firmly establish, or test for, any relations between the \nev\ line and any other AGN properties, and to pave the way for using \nev\ to study $z>6$ AGN with JWST, one has to obtain high-quality optical/near-IR spectroscopy for a large and complete sample of AGN for which such properties can be determined. The BAT AGN Spectroscopic Survey (BASS; \citealt{Koss2022_DR2_overview})\footnote{\url{http://www.bass-survey.com}} is acquiring optical spectroscopy, and many other multi-wavelength data, for a growing sample of (mostly) low-redshift AGN selected through their ultra-hard X-ray emission. This large dataset includes all types of AGN, including highly obscured systems, and benefits from having a variety of ancillary data products and measurements, particularly detailed X-ray spectral decomposition that provides the intrinsic hard X-ray luminosities and \NH\ \citep{Ricci2017_Xcat}; estimates of \mbh\ based on either broad emission lines and/or stellar velocity dispersions \citep{Koss2022_DR2_sigs,MejiaRestrepo2022_DR2_BLR, Caglar2023}; and various ways to estimate \Lbol\ \citep{Gupta24} and thus the Eddington ratio (\lamEdd\ hereafter).

In this paper we investigate the \NeV\ narrow line emission in a large subset of the BASS AGN, and look into any links between \nev\ emission and other AGN properties. Although the \NeV\ line was measured as part of the BASS DR2 spectroscopic catalog \citep{Oh2022_DR2_NLR}, we perform dedicated spectral fits and stacking in an attempt to maximize the detection rate and to remove some of the assumptions that had to be taken in the spectral fitting done in \citep{Oh2022_DR2_NLR}. We also note that another emission line originating from Ne$^{+4}$, namely \nev\,$\lambda\lambda$14.3,24.3 \mic, is the subject of a recently published, dedicated BASS study \citep{Bierschenk2024}, which we discuss throughout the present paper.
This paper is organized as follows. In Section~\ref{sec:obs_data}, we describe the sample and the spectroscopic data we use.
In Section~\ref{sec:analysis}, we describe our spectral analysis procedures (including fitting \& stacking). We present our findings regarding \nev\ line emission, and links with other AGN properties, and their implications in Section~\ref{sec:results}, and conclude in Section~\ref{sec:summary}.
Throughout this paper we assume a flat, cold dark matter cosmology with $H_0 = 70\,\Hubble$, $\Omega_\Lambda = 0.7$ and $\Omega_{\rm M}=0.3$. 
Unless noted otherwise, we use \nev\ to denote the (narrow component of the) \NeV\ emission line, and that line alone.\footnote{In our analysis, the narrow component is assumed to have $\fwhm<1200\,\kms$ (see Section~\ref{sec:spec_line}).}

%%%%%%%%%%%%%%%%%%%%%%%%%%%%%%%%%%%%%%%%%%%%%%%%%%%%%%%%%%%%%%%%
\section{Sample and Data}
\label{sec:obs_data}

\subsection{Sample}
\label{sec:sample}

Our sample is drawn from the 2nd data release of the BAT AGN Spectroscopic Survey (BASS/DR2; \citealt{Koss2022_DR2_overview}), which presented an unprecedentedly large and complete optical spectroscopic survey of AGN selected through their ultra-hard X-ray emission.
Specifically, BASS/DR2 collated optical spectroscopy, redshifts and basic spectral classifications for 858 AGN. This includes $>$95\% of all the AGN identified in the flux-limited 70-months Swift-BAT all-sky catalog \cite[beyond the Galactic plane, $|b|>5\deg$;][]{Baumgartner2013}.
In addition, BASS/DR2 provides spectra and basic properties for (X-ray) fainter AGN identified in deeper Swift/BAT data, down to the 105-month all-sky BAT flux limit ($8.2\times10^{-12}\,\ergcms$; \citealt{Oh18_105m}).

From this parent sample, we focus on narrow-line BASS/DR2 AGN, i.e. those classified as Seyfert galaxies of types $1.8-2$ in BASS/DR2 \cite[see][]{Koss2022_DR2_cat}. While \NeV\ emission is also observed in broad-line AGN, such systems are expected to have exceptionally weak \nev\ emission in terms of equivalent width ($\ewnev<1$ \AA; e.g., \citealt{VdB2001}), given the prominent, AGN-dominated blue continuum emission, which makes line measurements more challenging. In addition, the need to account for blended iron emission features and/or broad(er) \nev\ line profiles, requires particularly high $S/N$ data with high spectral resolution, which is not available for many of our BASS sources. Given the close physical link between the optical spectral classification and (X-ray) line-of-sight obscuration \cite[e.g.,][]{Oh2022_DR2_NLR}, this choice naturally leaves out most of the unobscured AGN in BASS (i.e., those with $\log[N_{\rm H}/\cmii]<21$).
There are, however, 14 sources with narrow emission lines but no signs of X-ray obscuration, i.e. with $\lognh=20$, nearly all of which are Seyfert 1.9 sources. Several studies have suggested that such systems may be explained by either particular viewing angles that obscure a significant fraction of the BLR but not the central X-ray source, or by a particularly low BLR covering factor in low-luminosity AGNs \cite[see, e.g.,][and references therein]{Barcons2003,Trippe2010,SternLaor2012,Burtscher2016}. In any case, the inclusion of a few low-\NH\ but narrow-lined sources in our sample does not affect our analysis and main results.

Next, to ensure high sensitivity and robust flux calibration towards the blue edge of the optical regime, we first narrow down our sample to include only sources for which blue optical spectroscopy (covering $\lambda<4000$\,\AA) was obtained either with the X-Shooter instrument at the VLT \citep{Vernet11} or the Double Spectrograph (DBSP) mounted on the 5 m Hale telescope at the Palomar observatory \citep{OG82}.
We finally omit from our sample any beamed AGN, as determined from their multi-wavelength data \citep{Paliya19,Marcotulli22}.
These selection steps leave us with a total of 432 spectra of narrow-line, non-beamed AGN, comprising of 190 X-Shooter and 242 DBSP spectra.

\begin{figure}
\centering
    \includegraphics[width=0.495\textwidth]{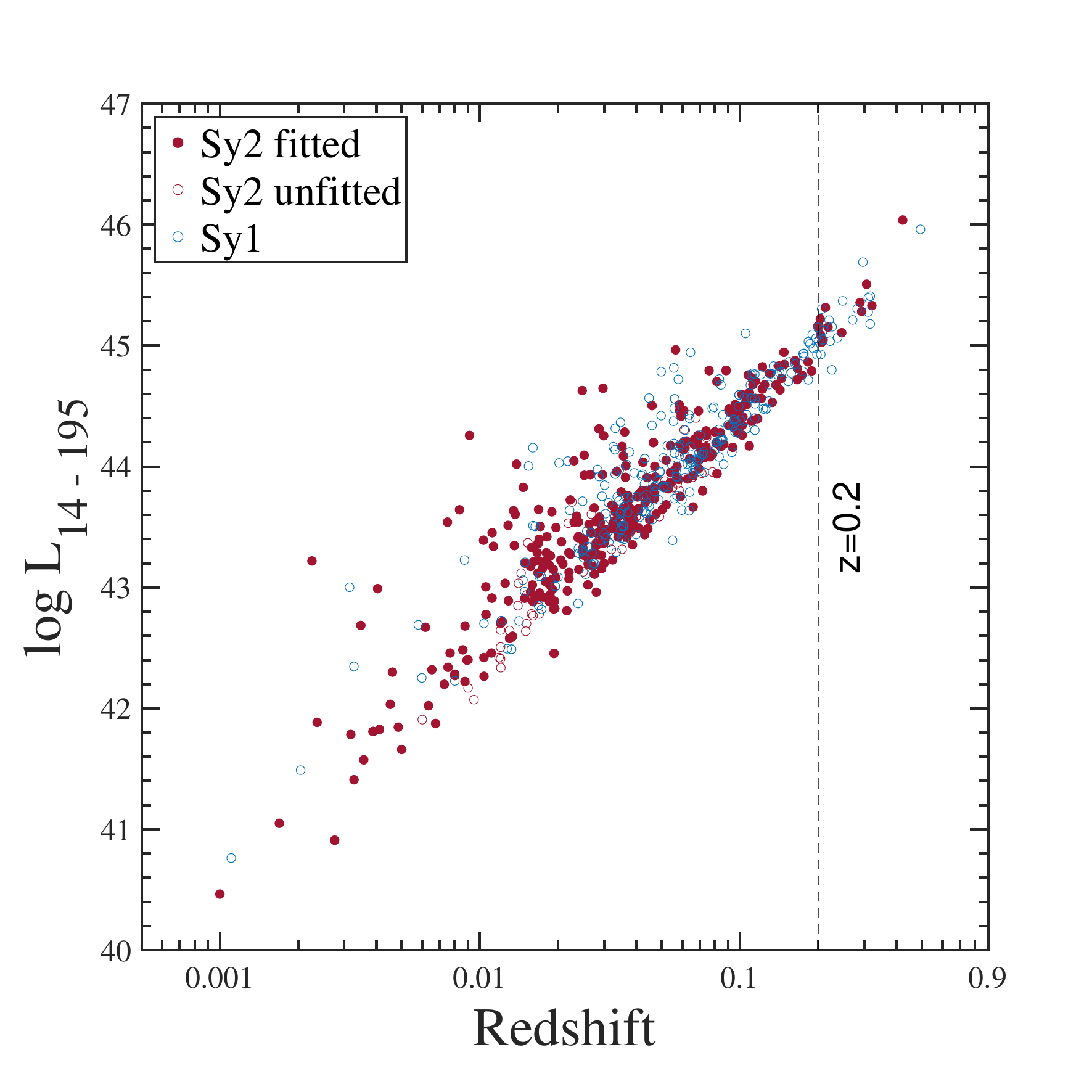}
    \caption{The luminosity-redshift plane for our sample and for BASS AGN in general. We show only those BASS/DR2 AGN with either VLT/X-Shooter or Palomar/DBSP spectra available. Red symbols mark narrow-line AGN, with filled circles further highlighting those AGN for which we derived useful measurements of, or constraints on, \NeV\ line emission. Empty blue symbols mark broad-line BASS/DR2 AGN with similar spectroscopic data, for comparison. Our sample lies mostly at $z\leq0.2$ ($\sim$97\% of sources; highlighted with a dashed vertical line), and is not biased compared with the general BASS/DR2 population.}
\label{fig:Lz}
\end{figure}

Figure~\ref{fig:Lz} shows our sample of BASS AGN in the redshift vs. intrinsic ultra-hard X-ray luminosity ($L[14-195\,\kev]$) plane, with the latter measurements taken from the \cite{Ricci2017_Xcat} catalog (see below). We further highlight the subset of \Nnev\ AGN for which our spectral analysis resulted in useful measurements of, or upper limits on, \nev\ emission (as explained in Section~\ref{sec:spec_line} below), as well as the broad-line BASS AGN for which similar spectroscopic data is available (i.e., X-Shooter or DBSP spectra), but which are not considered in the present work.
The vast majority of the sample considered here ($\sim$97\%; 417/432) are located at $z\leq0.2$. For $\sim$11\% of the narrow-line AGN in our initial sample (48/432), their DBSP spectra did not allow us to derive useful measurements, or even upper limits on, \nev\ emission, as the line is was very close to the noisy, blue end of the spectral coverage. 
All three subsets show the obvious bias resulting from the flux-limited nature of the Swift/BAT (and thus, BASS) data, with the lowest accessible luminosity increasing by nearly 3\,dex between $z=0.01-0.2$ (again - as expected from the change in luminosity distance). However, the key takeaway from Fig.~\ref{fig:Lz} is that our sample of narrow-line AGN with \nev\ measurements (or upper limits) well represents the parent BASS sample, and thus the population of bright, low-redshift, obscured AGN, and that our sample refinement does not introduce any additional biases. 

As explained in Section~\ref{sec:spec_line}, we performed detailed spectral measurements, followed by careful visual inspections, for all the remaining sources. After excluding several additional sources for which the data was not of adequate quality to even constrain the \nev\ line, and after removing duplicate spectra of a few AGN, we were left with \Nnev\ unique sources for which we could measure, or place upper limits on, \nev\ line emission.

We note that our sample covers a wide range of galactic latitudes and, as a result, a wide range of foreground attenuation ($E[B-V] \lesssim 3 $). While we do our best to account for this attenuation (see below), we stress that our main analysis and conclusions do not depend on whether highly foreground-attenuated sources (e.g., with $E[B-V]>0.2$) are included in, or excluded from, our sample.

\subsection{Spectroscopic Data}
\label{sec:spectra}

The X-Shooter and DBSP spectra we use here were obtained as part of the BASS efforts. These data are described in detail in \cite{Koss2022_DR2_overview,Koss2022_DR2_cat} and \cite{Oh2022_DR2_NLR}, and here we only mention the most basic properties of the data. 

The relevant part of the X-Shooter spectra, obtained in the UVB arm using a 1.6\arcsec slit, covers $\lambda=3000-5600$\,\AA\ with a sampling of 0.2\,\AA\ pix$^{-1}$ and a spectral resolution of $R\simeq3850$ \cite[see also][]{Koss2022_DR2_sigs}. The relevant part of the DBSP spectra, obtained in the blue arm with a 600/4000 grating, covers $3150-5600$\,\AA\ with 1.07\,\AA\ pix$^{-1}$ and $R=1220$.
In both sets of spectra, flux calibration was obtained using standard star observations, performed before and/or after the science observations. 

We stress that the quality of the spectra, in terms of their typical signal-to-noise ratio ($S/N$), varies significantly from one spectrum to another. 
This is due to the nature of the BASS spectroscopic observations, where some sources benefited from longer exposure times and/or preferable observing conditions, in a rather heterogeneous way (although exposure times were generally adjusted by broad-band target brightness).
We assess the effect of $S/N$ on our ability to detect \nev\ line emission below.

\subsection{Ancillary Data}
\label{sec:ancillary}

An important part of our analysis focuses on how \nev\ emission may vary with key AGN (or SMBH) properties. 
For this, we rely on the rich collection of measurements made available through the BASS project.

Specifically, we use intrinsic, obscuration corrected ultra-hard X-ray luminosities in the $14-150$\,\kev\ range (\Luhard\ hereafter), intrinsic, obscuration corrected hard X-ray luminosities in the $2-10$\,\kev\ range (\Lhard\ hereafter), and line-of-sight hydrogen column densities (\NH), as well as the associated uncertainties, that were determined through the exhaustive spectral decomposition of multi-mission X-ray data presented in \cite{Ricci2017_Xcat}. That study used a variety of spectral models, combining physical components in a way that best matches each BASS source. The typical uncertainties on these quantities are of order 0.1, 0.3, and 0.3\,dex, for \Luhard, \Lhard, and \NH\ (respectively).
Note that the BASS/DR2 catalog \cite{Koss2022_DR2_cat} includes some amendments and additions for a few sources, compared with the earlier catalog by \cite{Ricci2017_Xcat}. Still, we note that only \Nnh\ of the \Nnev\ AGN in the sample studied here have \NH\ estimates. 

While the Swift/BAT based selection of BASS strives to be complete even for highly obscured sources, the completeness rate naturally drops (see, e.g., Fig.~1 of \citealt{Ricci2015_CT}), and the uncertainty on derived X-ray properties grows, with increasing \NH. We should therefore be cautious about any findings that may be obtained for sources with $\lognh\gtrsim24.5$, although BASS still likely provides one of the most complete and best studied sample of AGN in this regime.

Black hole mass estimates are available for \Nmbh\ of the \Nnev\ AGN in our sample, through the dedicated BASS/DR2 catalog by \cite{Koss2022_DR2_sigs}. That work measured the stellar velocity dispersions (\sigs) in the hosts of narrow-line BASS AGN by fitting several stellar absorption features, and then used the well-known $\mbh-\sigs$ scaling relation to yield \mbh\ estimates (specifically, following the scaling derived by \citealt{KH13}). While the \sigs\ measurements are quite precise (and accurate), with measurement uncertainties of $\lesssim$0.1\,dex, the resulting \mbh\ estimates are dominated by systematic uncertainties and intrinsic scatter that may reach $\approx$0.4\,dex (see, e.g., \citealt{KH13,Koss2022_DR2_sigs,Caglar2023} and references therein, for a detailed discussion).

Finally, Eddington ratios (\lamEdd) were derived from available estimates of \mbh\ and of the bolometric luminosities of our AGN (\Lbol), where for the latter we rely on \Luhard, and assume a universal bolometric correction corresponding to $\Lbol= 8 \times \Luhard$ (consistent with other BASS/DR2-based studies; see discussion in \citealt{Koss2022_DR2_cat}). Assuming a solar metallicity composition for the gas around the SMBH, we obtain $\lamEdd = (\Lbol/\ergs) / [1.5\times10^{38}\,(\mbh/\Msol)]$.
In addition to the aforementioned uncertainties on \mbh, our \lamEdd\ estimates are affected by the uncertainty on the bolometric correction, which may exceed 0.5\,dex \cite[e.g.,][]{Gupta24}. The combined systematic uncertainty on \lamEdd\ may thus exceed 0.7\,dex. Moreover, AGN bolometric corrections were suggested to vary with AGN luminosity, \lamEdd\ itself, and perhaps other AGN properties \cite[e.g.,][]{Marconi04,Vasudevan07,Duras20,Gupta24}. Given that our analysis did not yield any significant correlations between \nev\ emission and \lamEdd, we prefer to use the aforementioned single, universal bolometric correction, for the sake of simplicity. We verified that alternative choices of bolometric corrections would not have affected our main conclusions, and in particular would not have resulted in a statistically significant (anti-)correlation between (relative) \nev\ strength and \lamEdd.

%%%%%%%%%%%%%%%%%%%%%%%%%%%%%%%%%%%%%%%%%%%%%%%%%%%%%%%%%%%%%%%%
\bigskip\bigskip\bigskip\bigskip
\section{Spectral Analysis}
\label{sec:analysis}

\subsection{Line Measurements}
\label{sec:spec_line}

To measure, or place robust constraints on, narrow \nev\ line emission in our (sub-)sample of BASS AGN, we decomposed and modeled the relevant spectral region as explained below.

We first de-redshifted all spectra to the rest-frame, using the redshifts provided in the BASS/DR2 catalog, which are based on prominent narrow emission lines (e.g., \oiii; \citealt{Oh2022_DR2_NLR}).
For each spectrum we then fit a simple, linear continuum model to the observed flux densities (i.e., $F_\lambda = m \lambda + b$) in two 14\,\AA-wide windows, separated by $\pm1200\,\kms$ ($\pm2000\,\kms$) from the expected location of the \nev\ line\footnote{We assume a central wavelength of 3425.88\,\AA\ in air.} in the X-Shooter (DBSP) spectra, that is rest-frame wavelengths of $3405-3419$ \& $3433-3447$\,\AA\ for the X-Shooter spectra and $3396-3410$ \& $3442-3456$\,\AA\ for the DBSP ones. The linear continuum model is then subtracted from the spectrum.

We next model the \nev\ emission line in the continuum-subtracted spectra with a single Gaussian profile. As we are only interested in narrow \nev\ emission, the Gaussian width is constrained to have full width at half maximum of ${\rm FWHM} = 120-1200\,\kms$ or $200-1200\,\kms$, for the X-Shooter or DBSP spectra, respectively. The lower bounds on FWHM were chosen so they would always correspond to $\gtrsim1.5\times$ the spectral resolution of the corresponding spectra. Our visual inspection (see below) suggests these lower bounds did not bias the model fitting. The Gaussian center is allowed to shift by up to 360\,\kms\ relative to the expected central wavelength of the line, to account for minor issues with redshift determination, wavelength calibration, or the potential effects of ionized gas outflows. The fitting is performed using a standard $\chi^2$ minimizing algorithm.

We visually inspected all the best-fitted models to identify catastrophic failures and to refine some of the aforementioned parameter choices. We note that in several cases among our BASS AGN the \nev\ line profile showed blue wings and/or shoulders, suggestive of outflowing gas photoionized by the AGN itself. We show one such case in Appendix~\ref{app:outflows}, but any further discussion of this phenomena is beyond the scope of the present study.

The fitted line and continuum models can be used to derive the total line flux (\fnev) and rest-frame equivalent width (\ewnev) of the \nev\ line. The uncertainties on these key \nev\ measurements are estimated through a re-sampling approach. For each AGN, we produce 100 mock spectra by jittering the observed flux density, at each spectral pixel, by an offset that is drawn from a normal distribution with a width equal to the error on the observed flux density at that spectral pixel. 
Each of the 100 mock spectra is fitted following the same procedure as the original spectrum, providing a distribution of \fnev\ and \ewnev\ for every AGN. We stress that in this re-sampling and re-fitting procedure, all model parameters (including line width and shift) are free to vary, as in the fitting of the original spectra. 
The 0.16 and 0.84 quantiles of the \fnev\ and \ewnev\ distributions are then used as the measurement uncertainties on these key emission line quantities.\footnote{Our fitting and re-sampling procedures also provide measurements, and associated uncertainties, on the width and shift of \nev, however these are not used in the analysis that follows.}

Throughout our analysis, we use a combination of criteria  to identify spectra in which a narrow \nev\ emission line is robustly detected. The most important and simplest criterion is $\fnev/\Delta\fnev > 3$, assuring that the line emission is statistically significant. We additionally require that the continuum emission is also robustly detected, with $S/N>3$ in the continuum bands. Given the brightness of our sources and the somewhat heterogeneous nature of our spectra (in terms of depth), this criterion is meant to assure that the data is of sufficient quality to make weak line measurements, and specifically make the \ewnev\ measurements robust.
We further required that the line width measurements satisfy $\fwhm(\nev)-\Delta\fwhm(\nev)\leq1198\,\kms$. This criterion assures that the best-fitting line profile model indeed captures the {\it narrow} \nev\ line, since for broad \nev\ profiles the distribution of FWHM derived from our line-fitting procedure ``saturated'' at the 1200\,\kms\ boundary (i.e., a best-fitting value of 1200\,\kms\ with negligible uncertainties).
Our extensive visual inspection suggests that these criteria are mostly adequate in capturing the (wide) range of spectral data and fit quality for our large sample.
During our visual inspection, we did flag a handful of additional sources as lacking robust \nev\ detection and measurements despite satisfying all these criteria. This was usually due to catastrophic issues with the relevant parts of the spectrum and/or peculiar line profiles where a narrow line core cannot be easily identified.
For the sources that don't qualify as \nev-detected, we use $3\sigma$-equivalent upper limits on \fnev\ and---combined with the best-fitting continuum models---on \ewnev. 
After applying all quality checks and detection criteria, we have measurements or upper limits on \fnev\ (and on \ewnev) for a total of \Nnev\ unique narrow-line AGN.

The \nev\ flux measurements and upper limits were corrected for foreground (Milky Way) extinction using the maps of \cite{Schlegel1998} and assuming the \cite{Cardelli1989} extinction law, with $R_{\rm V}=3.1$. 
We note that, being a relatively short-wavelength (optical) emission line, \nev\ is also affected by extinction in the host galaxies of our AGN. However, we choose not to apply any further corrections to the fluxes, given the great uncertainties in any assumption on (inner) host-scale extinction, even if information about the stellar/gas content, morphologies, and orientations of the hosts is available (in practice, such information is heterogeneous and not yet complete for the BASS sample).

\subsection{Spectral Stacking}
\label{sec:spec_stack}

Since \nev\ is known to be weak even in powerful AGN, and since one of our goals is to search for links between \nev\ emission and other AGN properties, we have also constructed stacked \nev\ spectra for various subsets of AGN, which share common properties. The stacking was performed separately for the VLT/X-Shooter and Palomar/DBSP subsets of spectra, given the differences in their spectral sampling, resolution, and overall quality. Specifically, we stacked spectra of AGN in bins of \Luhard, \NH, \mbh, and \lamEdd. The bin sizes were chosen to reflect the different number of AGN in the two spectroscopic datasets, and the varying number of AGN across the range of the  properties we examined. For the VLT/X-Shooter spectra, all bins were 0.5\,dex wide, except for \NH\ where we used 1\,dex wide bins. For the smaller Palomar/DBSP dataset, we used 1\,dex wide bins also for \Luhard\ and \lamEdd. We produced stacks from either all AGN in our sample, or---alternatively---only from those AGN where \nev\ line emission was \textit{not} detected (i.e., individually). In Section~\ref{sec:stack_res}, however, we choose to focus only on the former set of stacks. With these choices, the number of spectra per bin varied between as few as three and as many as 47. The various bins and the number of spectra in each of them, for each of the spectral subsets, are listed in a dedicated Table in Appendix~\ref{app:stack_tab}.

The spectral stacking itself was performed as follows.
For every AGN belonging to a particular subset (facility and bin), we first fit a linear continuum model, as described above, and then divide the spectrum by the continuum model, such that the flux density in the line-free regions is $F_{\rm \lambda, norm} = 1$ (on average). The continuum-normalized spectrum is then re-sampled to a fixed linear wavelength grid with either $\Delta \lambda = 0.25$ or 0.5\,\AA\,pix$^{-1}$, for X-Shooter or DBSP, respectively. All the continuum-normalized spectra belonging to a given subset are then combined to form the composite spectra representative of that subset, by taking either the average or median values, per (uniform) spectral pixel.\footnote{We have also experimented with constructing composite spectra based on geometrical means, but found them to carry no additional information.}
We modeled the \nev\ emission line in the composite spectra in the same way as for the individual BASS spectra, focusing on the resulting \ewnev\ (given that \fnev\ has no physical meaning in these normalized, composite spectra).

Since the uncertainties on the composite spectra are dominated by the (limited) statistics of the individual AGN rather than on measurement uncertainties, the corresponding uncertainties on \ewnev\ were calculated through a Jackknife approach. Specifically, for each subset of spectra, we repeated the stacking procedure 100 times, each time re-drawing a random selection of the spectra belonging to that subset (with repetitions). The \nev\ emission in each of the resulting composites was modeled and the resulting distributions of \ewnev\ were used to derive $\Delta\ewnev$ (again using the 0.16 and 0.84 quantiles). 

Here, too, we adopt a the aforementioned criteria to identify those stacks in which \nev\ emission is robustly identified. 
A visual inspection of the stacks suggests that our criteria represent a somewhat conservative choice: in some cases, the \nev\ emission line seems to appear in the stacked spectrum, however the uncertainty on this--which is dominated by the statistics of selecting AGN for each bin--is too large to deem the measurement robust.
We particularly reiterate the requirement of continuum $S/N>3$, and of limiting our analysis to only {\it narrow} \nev\ line emission, as some of the stacks present apparently strong \nev\ signal over a noisy (stacked) continuum level, and/or a broad (stacked) \nev\ profile.
We return to this point when discussing the stacking results, in Section~\ref{sec:stack_res}.

%%%%%%%%%%%%%%%%%%%%%%%%%%%%%%%%%%%%%%%%%%%%%%%%%%%%%%%%%%%%%%%%
\section{Results and Discussion}
\label{sec:results}

The present study focuses on what could be inferred from those AGN in which \nev\ is detected, and indeed from the observed line strength. We therefore note that in what follows we chose to:
\begin{itemize}
    \item Not make any attempt to correct the \nev\ emission for host galaxy extinction.
    \item Focus on scaling relations, and correlation tests, that involve \nev-detected sources.
\end{itemize}
These choices are further motivated by our desire to provide scaling relations that can be used ``as-is'' for AGN with much more limited data than what is available for BASS AGN, particularly high-redshift systems observed with limited spectral coverage and/or ancillary (multi-wavelength) data.

\subsection{Detection Fraction}
\label{sec:detect_frac}

Following our quantitative criteria and the further visual inspection, we find that significant narrow \nev\ line emission is detected in \Ndet\ of the \Nnev\ AGN in our final sample,  i.e. an overall detection fraction of $\fdet=42.8\%$. We note that this may be considered a rather conservative lower limit, given the multiple criteria we imposed (i.e., continuum $S/N$; see Section~\ref{sec:spec_line} above).

\begin{figure}
    \centering
    \includegraphics[clip, trim=6.5cm 1.0cm 3.5cm 0.1cm,width=0.5\textwidth]{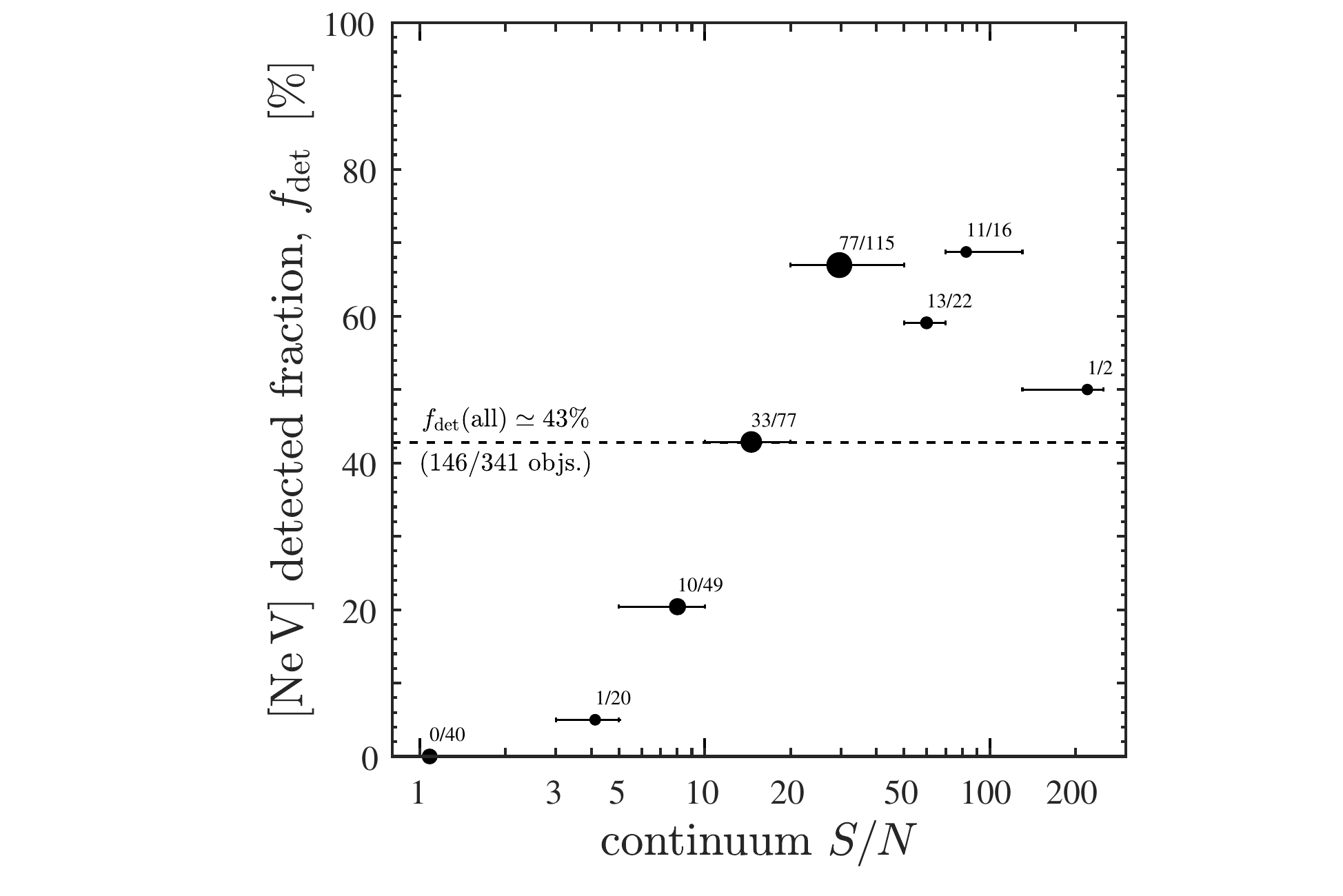}
    \caption{The detection fraction of \NeV\ and its dependence on data quality. We group our BASS spectra into several bins in $S/N$ (indicated along the horizontal axis; mind the logarithmic scaling), and calculate the percentage of spectra where \NeV\ was robustly detected. The corresponding numbers of objects are indicated next to each data point, and the size of the data points scales with the (total) number of objects in each bin. The horizontal error-bars indicate the range of $S/N$ for the spectra included in each bin. The horizontal dashed line marks the overall \nev\ detection fraction among our entire sample, $\fdet\simeq43\%$. Even in high $S/N$ data, the detection fraction does not exceed $\sim70\%$.} 
    \label{fig:detper}
\end{figure}

One may naturally expect that our ability to detect \nev\ would depend on the quality of the spectra in hand. Figure~\ref{fig:detper} shows \fdet\ in several bins of continuum $S/N$. While there is indeed a trend of decreasing \fdet\ towards the lowest (acceptable) $S/N$, we note that at higher $S/N$ the trend saturates near $\fdet\simeq75\%$ at $S/N\gtrsim30$. Given that in practice this is an exceptionally high $S/N$, which is very rarely reached in spectroscopic campaigns of larger and/or higher redshift samples, we conclude that the intrinsic \fdet\ should, in all likelihood, never reach $\simeq100\%$.

In Section~\ref{sec:props_trends} we look into potential links between the (relative) strength of \nev\ emission and key AGN properties, specifically \Luhard, \NH, \mbh, and \lamEdd. Concerning the detection fraction, on top of each of the panels of the relevant Figures~\ref{fig:nev_vs_props_1} and \ref{fig:nev_vs_props_2} we show how \fdet\ varies across the range of these quantities covered by our sample. We generally see no strong trends between \fdet\ and any of these properties. We note in particular that the detection fraction remains high, $\fdet\approx60\%$, even at $\lognh \gtrsim 23$. Similarly, \fdet\ exceeds $\gtrsim35\%$ both at the lowest and highest ends of the \lamEdd\ range of our sample ($\log\lamEdd\simeq -2.75$ and $0.5$, respectively).

We conclude that \nev\ is robustly detected in a significant fraction, but definitely not all, of our complete sample of ultra-hard X-ray selected AGN, and that this fraction is largely independent of other key AGN/SMBH properties. The detection rate remains well below $100\%$ even for luminous sources with extremely deep spectroscopy and no signs of extremely high line-of-sight obscuration. We discuss some possible explanations for this in Section~\ref{sec:dis_low_nev}.

\subsection{Scaling with X-ray emission}
\label{sec:LL_scaling}

\begin{figure*}
\centering
    \includegraphics[clip, trim=5.25cm 0.5cm 5.6cm 0.25cm,height=0.425\textwidth]{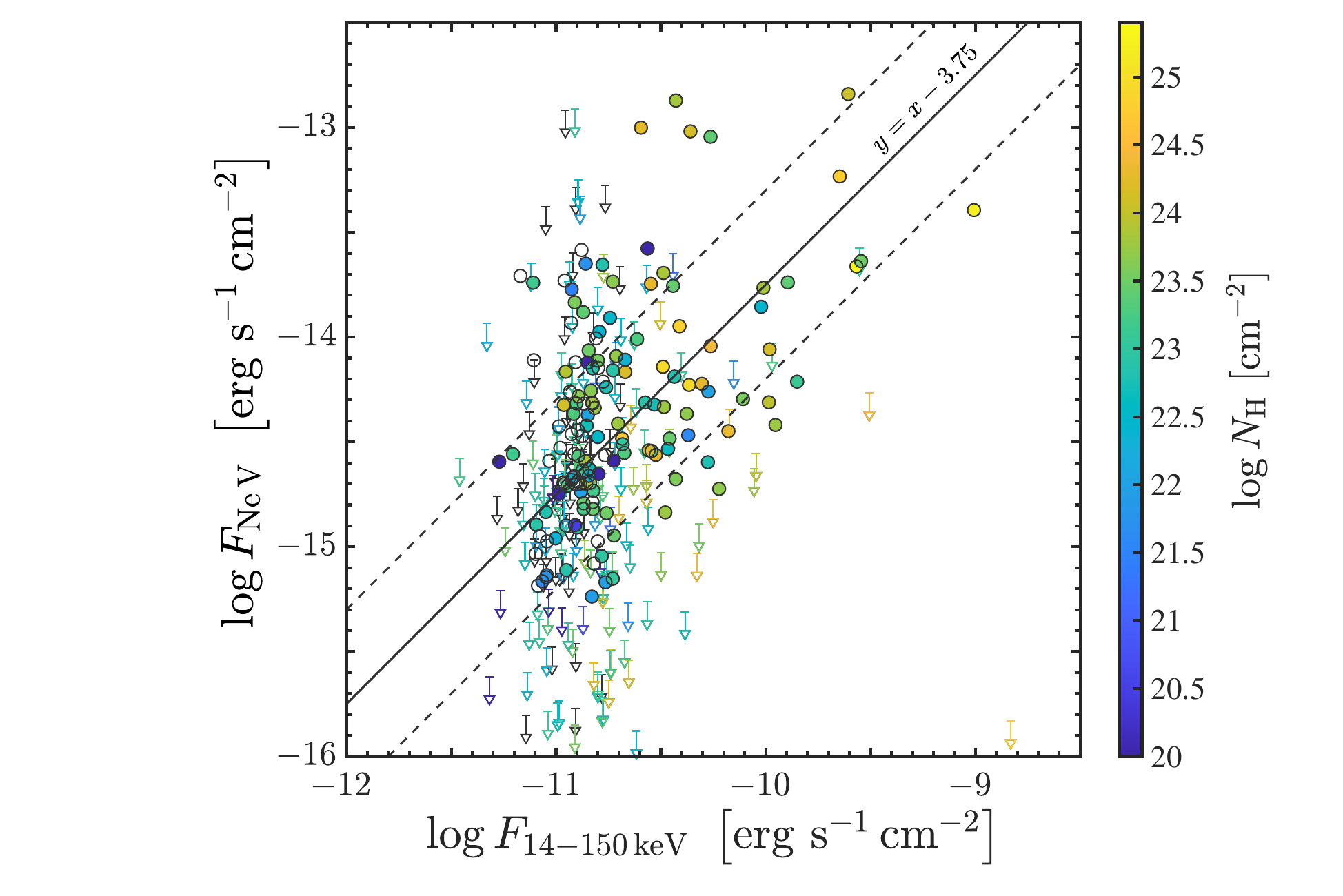}\hfill
    \includegraphics[clip, trim=4.25cm 0.5cm 1.25cm 0.25cm,height=0.425\textwidth]{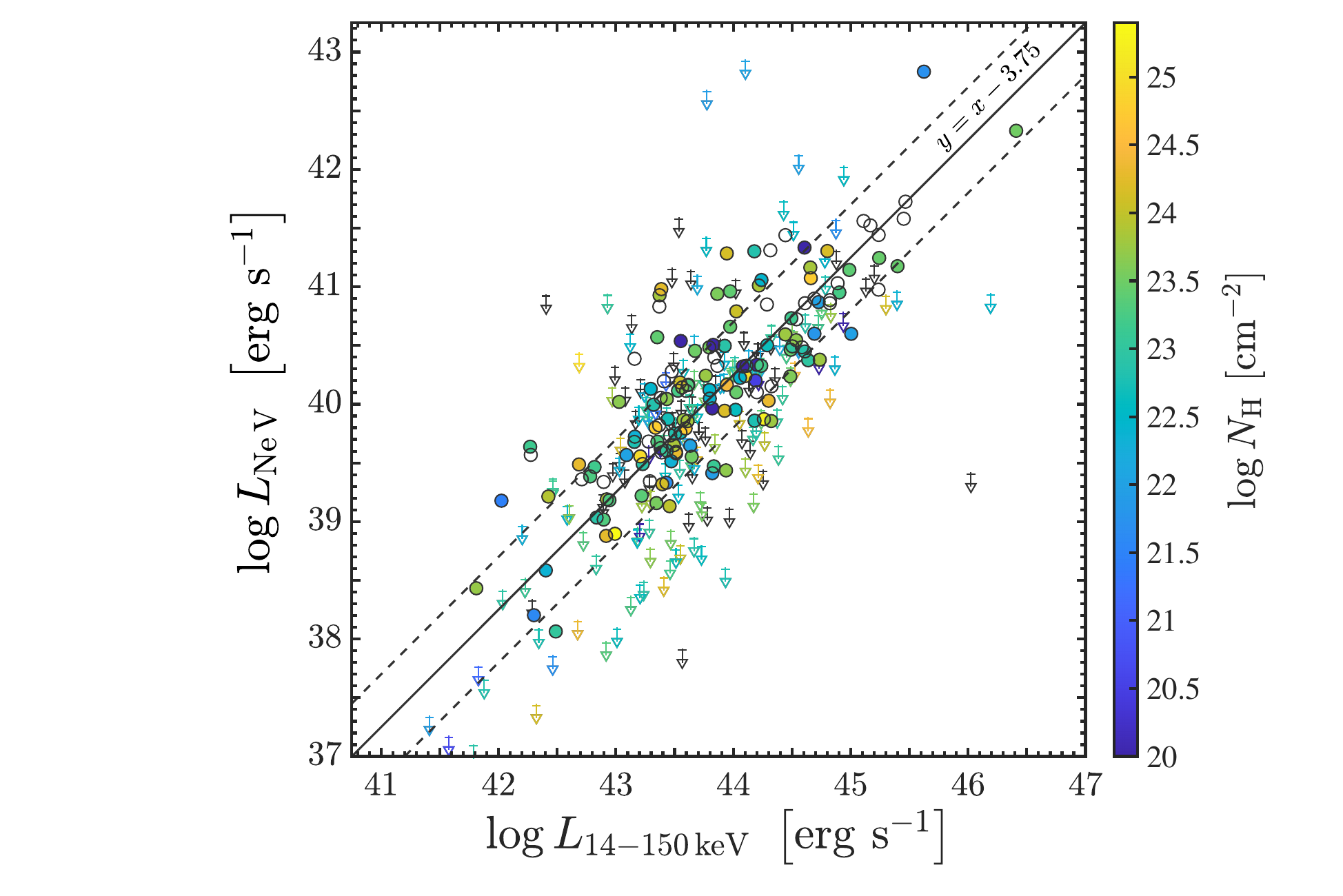}
    \caption{The relation between \NeV\ and ultra-hard X-ray emission for our BASS sample of AGN.
    We show both the flux-flux (\fnev\ vs.\ \Fuhard; \textit{Left}) and luminosity-luminosity (\Lnev\ vs. \Luhard; \textit{right}) scatter plots.
    In both panels, the \nev\ measurements (filled circles) and upper limits (downward facing arrows) are color-coded based on \lognh\ (see vertical colormap). The solid dashed black lines mark a linear relation scaled to match the median ratio between \nev\ and $14-150$\,kev emission, while dashed lines run parallel, with $\pm$0.3\,dex offsets.} 
\label{fig:nev_vs_bat}
\end{figure*}

We next look into the relation, and scaling, between narrow \NeV\ line emission and the (ultra) hard X-ray emission of our AGN. Figure~\ref{fig:nev_vs_bat} presents the relation between integrated \nev\ and $14-150$\,kev emission strength, in terms of both fluxes (left) and luminosities (right), with markers color-coded by \NH. There are two key points evident in Fig.~\ref{fig:nev_vs_bat}. First, \nev\ emission is robustly detected across a wide range of AGN luminosities (as traced by the ultra-hard X-ray emission), and across all levels of nuclear obscuration (as traced by \NH), up to and including Compton-thick levels ($\log[N_{\rm H}/\cmii]\gtrsim24$). We look more closely into trends with \NH\ in Section~\ref{sec:props_trends} below. Second, there's a clear correlation between these markedly different emission components, albeit with considerable scatter. We confirm the statistical significance of the flux-flux correlation (left panel) through appropriate correlation tests, finding $P<10^{-5}$ for both the Spearman and Pearson correlation tests employed for the \nev\ measurements, and $P_{\rm K}<10^{-5}$ for the Kendall $\tau$-test, which also accounts for upper limits on \fnev.
A simple least-squares fit to the luminosity measurements of the \nev-detected sources (right panel) yields the relation:
\begin{equation}
   \label{eq:Lnev_Luhard}
    \begin{split}
    \log\left(\frac{\Lnev}{10^{40}\,\ergs}\right) = (0.86\pm0.12)\times\log\left(\frac{\Luhard}{10^{44}\,\ergs}\right) \\ 
    + (0.30\pm0.09) \, .
    \end{split}
\end{equation}
The scatter (standard deviation) of the residuals around this best-fitting luminosity-luminosity relation is $\approx$0.3\,dex.

\begin{figure*}
\centering
    \includegraphics[width=0.5\textwidth]{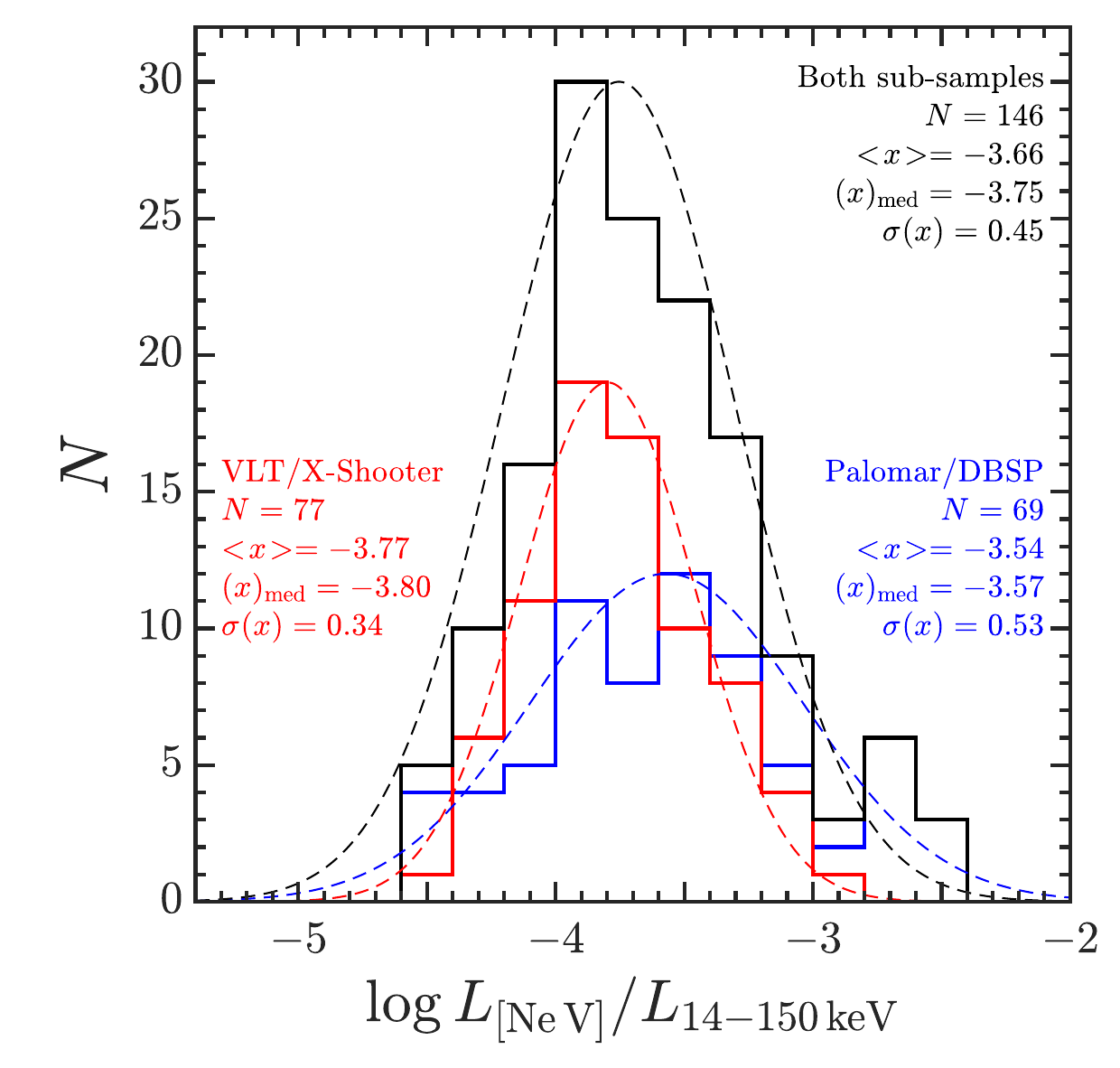}\hfill
    \includegraphics[width=0.5\textwidth]{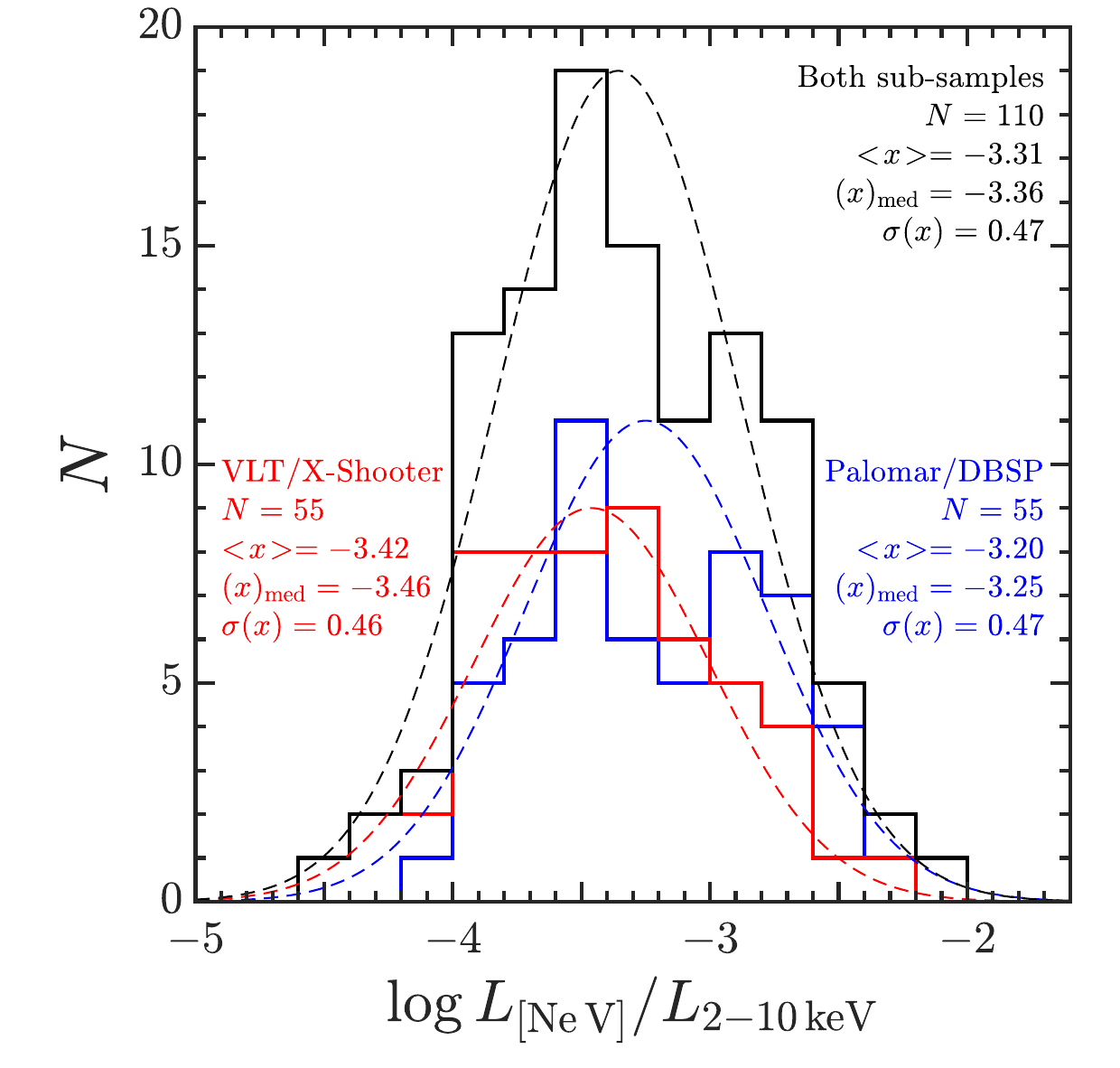}
    \caption{The ratio between \NeV\ and X-ray emission for the \nev-detected sources among our sample of BASS AGN. We show distributions of both $\Lnev/\Luhard$ ({\it left}) and $\Lnev/\Lhard$ ({\it right}). In both cases, we show the distributions derived from \nev\ measurements (solid lines) and upper limits (dashed lines).
    We also plot (scaled) normal distributions that have the same mean and standard deviation values as do the corresponding observed distributions (measurements of the entire sample).}
\label{fig:nev_xray_ratio}
\end{figure*}

The left panel of Figure~\ref{fig:nev_xray_ratio} shows the distribution of $\Lnev/\Luhard$ among the \nev-detected AGN in our sample, with various lines tracing \nev\ measurements for our entire BASS sample, and the separate X-Shooter and DBSP subsets. For any of these datasets, the distribution of $\Lnev/\Luhard$ is clearly unimodal and appears rather symmetric, bearing in mind that the low-$\Lnev/\Luhard$ end of the distribution may extend into the regime where \nev\ is not detected in our spectra. To emphasize this, we also show normal distributions (i.e., Gaussians), which have the same medians and standard deviations as do the corresponding distributions of $\Lnev/\Luhard$ measurements. For our entire BASS-based sample we find that the mean of the (logarithm of the) ratio is $\left<\log(\Lnev/\Luhard)\right> = -3.64$ and the median value is $\log(\Lnev/\Luhard)_{\rm med} = -3.75$, while the standard deviation is $\sigma[\log(\Lnev/\Luhard)] = 0.45$\,dex. We note that for our higher-quality X-Shooter sample the distribution of $\Lnev/\Luhard$ among \nev-detected sources is even tighter, with $\sigma[\log(\Lnev/\Luhard)]=0.34$\,dex.

As most X-ray surveys of (high-redshift) AGN are conducted at lower (softer) energies than those probed by Swift/BAT (i.e., with Chandra or XMM-Newton), we further looked at the scaling of \nev\ emission with rest-frame $2-10$\,\kev\ emission. The corresponding intrinsic luminosities, \Lhard, were taken from the \cite{Ricci2017_Xcat} BASS catalog of X-ray properties, where they were derived from the same multi-mission X-ray datasets, and the same spectral models, used to derive \Luhard. The right panel of Figure~\ref{fig:nev_xray_ratio} shows the distribution of $\Lnev/\Lhard$ for the \nev-detected AGN in our sample. Compared with the $\Lnev/\Luhard$ distribution, this distribution appear to have a larger scatter, and the unimodal nature is not as clear. The corresponding mean ratio we find in this case, among \nev-detected sources, is $\left<\log(\Lnev/\Lhard)\right>=-3.31$, while the corresponding median is $\log(\Lnev/\Lhard)_{\rm med}=-3.36$ and the standard deviation is $\sigma[\log(\Lnev/\Lhard)]=0.47$\,dex.
In this case, the scatter in the distribution of the higher quality X-Shooter measurements is not notably smaller than that of the entire sample ($\sigma[\log(\Lnev/\Luhard)]=0.46$ vs.\ 0.47, respectively), but the distribution itself appears to be more symmetric than that of the DBSP subset.

Several previous works studied the links between emission lines of highly-ionized species, including \nev, and (ultra) hard X-ray emission.

Focusing first on \nev\ transitions, studies that used Spitzer/IRS spectroscopy focused on the two MIR lines of \nev\ at 14.32 and 24.32 \mic. The study by \cite{Weaver2010} focused on 79 AGN drawn from earlier phases of the Swift/BAT survey for which high-resolution Spitzer/IRS spectroscopy allowed the detection of at least one of the MIR \nev\ lines for $>$90\% of the sources. They found remarkably strong and tight correlations between \nev\ MIR line and ultra-hard X-ray luminosities, for both obscured and unobscured sources, however the total range in the $\Lnev({\rm MIR})/\Lbat$ ratio was found to be $\approx$2\,dex.
The most recent study of MIR \nev\ emission among Swift/BAT selected AGN was presented by the BASS team in \cite{Bierschenk2024}. The analysis was based on Spitzer/IRS spectra collated for 140 of the BASS AGN, again including both obscured and unobscured sources, and harnessing the ancillary data available through BASS/DR2. \cite{Bierschenk2024} also found a strong and tight correlation between \nev\ and ultra-hard X-ray luminosities. The correlations found in \citet[their Fig.~5]{Bierschenk2024} appear tighter than what we find here, but in fact exhibit a level of scatter that is highly consistent with our findings. Specifically, they report a scatter of 0.5\,dex between $L_{\nev\ 14.3\,\mic}$ and \Luhard, and 0.8\,dex between $L_{\nev\ 14.3 \mic}$ and \Lhard.
Taken at face value, the high detection rate of the MIR \nev\ lines reported in \cite{Bierschenk2024}, of $\approx80\%$, may suggest that the \NeV\ line studied here could be affected by (host-scale) dust attenuation, given that the MIR lines are naturally unaffected by dust. The comparable scatter in the \nev-to-X-ray relations, however, supports the notion that such host-scale dust attenuation does \emph{not} strongly affect our analysis and findings. 
We further discuss the possible effects of host-scale dust on the detection rate and relative strength of the \NeV\ line in Section~\ref{sec:dis_low_nev} below.
Another recent study by \cite{Spinoglio2022} found very similar results for a sample of 100 AGN from the complete 12 \mic\ sample of Seyfert galaxies with MIR spectroscopy. Focusing specifically on highly obscured, Compton-thick AGN, \cite{Spinoglio2022} clearly show that such sources do not differ from other AGN in terms of detection rate of their MIR \nev\ lines or their positions in the \nev-to-X-ray parameter space(s). This is in good agreement with what we see in Figure~\ref{fig:nev_vs_bat}, where highly obscured sources (see colorbar) do not seem to occupy any particular region in the \nev-to-X-ray parameter space.

For completeness, we also note that some studies investigated the links between MIR \nev\ line emission and optical AGN continuum emission, for unobscured sources \cite[e.g.,][]{Dasyra2008}, with key results that are consistent with those focusing on X-ray AGN continuum emission (i.e., strong correlations with a scatter of $\sim$0.5\,dex).
Regardless of the strength (or tightness) of any links between the MIR \nev\ lines and intrinsic AGN emission (as traced by ultra-hard X-ray emission), we note that the MIR lines can only be accessed in the low-redshift Universe, and therefore their utility as probes of high-redshift AGN is extremely limited.

The high-ionization emission line that is perhaps the most commonly accessible and used to study AGN is \OIII. The work of \cite{Heckman05} looked into the ratio between \OIII\ and hard X-ray emission for $\lesssim100$ low-redshift AGN. Of these, the sample most comparable to ours consists of 47 X-ray selected AGN (both narrow- and broad-line). The commonly used \oiii-to-X-ray scaling reported in that study has a scatter of 0.5\,dex. Moreover, the subset of \oiii-selected, narrow-line AGN among the \cite{Heckman05} sample shows significantly larger scatter, skewed towards X-ray weak sources. A tight, nearly linear relation between $\log \Loiii$ and $\log\Lhard$ was also reported by \cite{Panessa06}. These \oiii-to-X-ray relations have been extensively used to determine the radiative outputs, and thus BH accretion rates, of several large samples of AGN, extending beyond the nearby Universe \cite[e.g.,][]{Hopkins07,Silverman09}.
More recently, the studies of \cite{Ueda2015_OIII} and \cite{Berney2015} found a scatter of $\approx$0.6\,dex in both $\Loiii/\Luhard$ and $\Loiii/\Lhard$ among large samples of nearby, ultra-hard X-ray selected AGN drawn from BASS/DR1. The scatter in these ratios among the subsets of Type 2 BASS/DR1 AGN was found to be even higher, approaching $\approx$0.7\,dex. \footnote{\cite{Ueda2015_OIII} further demonstrated that accounting for host-scale dust attenuation would result in yet larger scatter in the \oiii-to-X-ray ratio(s).}

Given these previous findings for \oiii, the scatter we find for the $\Lnev/\Luhard$ ratio among the \nev-detected AGN in our BASS sample, and particularly among the higher quality X-Shooter dataset, shows that---when detected---the \NeV\ line could be used as a reliable tracer of AGN radiative output, most likely even more than \oiii. This is not surprising, given that \oiii\ emission is contaminated by ionized gas in SF regions, while \nev\ is driven (almost) solely by the central AGN. We note that there's a non-negligible level of uncertainty when further converting (ultra-hard) X-ray luminosities to bolometric AGN luminosities \cite[e.g.,][and references therein]{Duras20,Gupta24}, which should be considered whenever one would like to derive (rough) estimates of \Lbol\ for any AGN (sample) with partial spectral coverage.

\subsection{Trends with other AGN properties}
\label{sec:props_trends}

\begin{figure*}
\centering
    \includegraphics[clip, trim=1.5cm 0.0cm 0.0cm 0.1cm, width=0.75\textwidth]{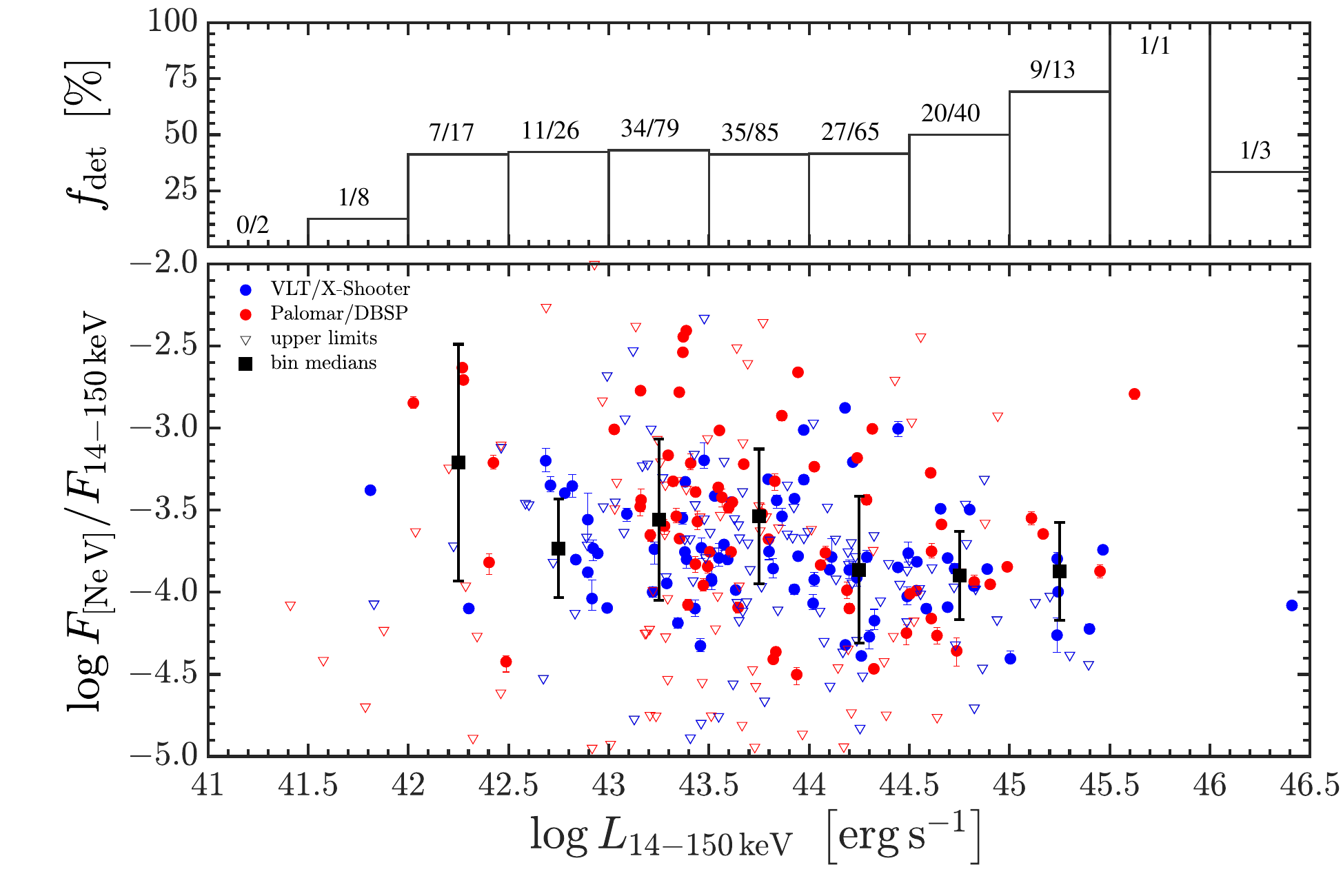}\\
    \includegraphics[clip, trim=1.5cm 0.5cm 0.0cm 0.0cm, width=0.75\textwidth]{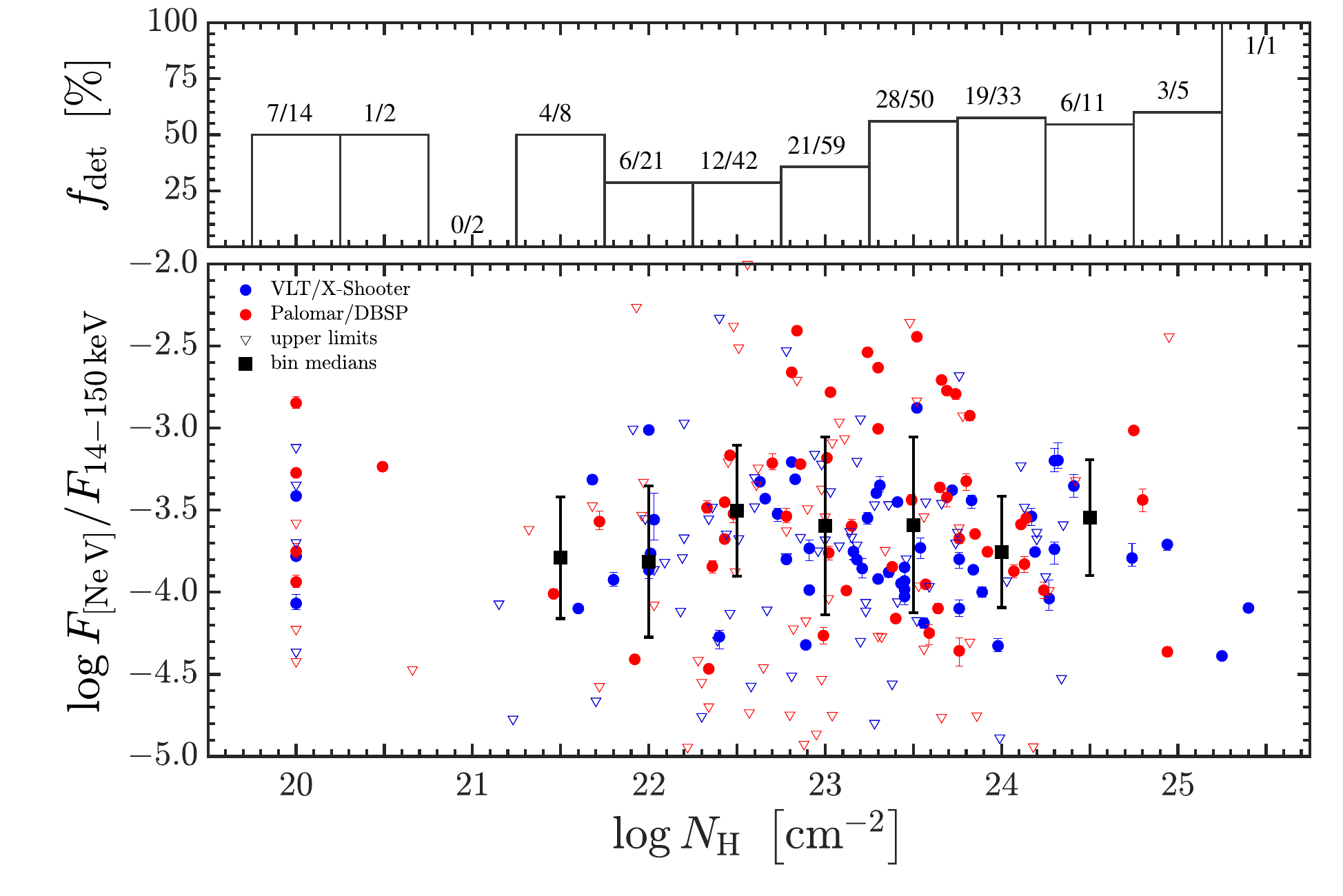}    
    \caption{The relation between relative \nev\ line strength, \ffx, and key properties of our BASS sample deduced from X-ray observations, including ultra-hard X-ray luminosity (\Luhard, top), and line-of-sight hydrogen column density (\NH, bottom). For each property, the various symbols mark measurements and upper limits derived from X-Shooter and DBSP spectra, as indicated in the legend. Large black symbols mark the median values among the \nev-detected sources, in running bins of the respective property, with error-bars indicating the standard deviations. In both cases, we see a significant scatter and no evidence for strong correlations, however the mild anti-correlation between \ffx\ and \Luhard\ is statistically significant, and echoes the sub-linear relation between \Lnev\ and \Luhard\ (see Eq.~\ref{eq:Lnev_Luhard}). The smaller panels on top of each scatter plot show how the \nev\ detection fraction, \fdet, varies across the range of the respective property. There are no strong trends in \fdet\ across the range of properties shown. Note in particular the persistent detection of \nev\ emission even in the highest column densities probed, $\lognh\gtrsim24$ (bottom panel).}
\label{fig:nev_vs_props_1}
\end{figure*}

\begin{figure*}
\centering
    \includegraphics[clip, trim=1.5cm 0.0cm 0.0cm 0.1cm, width=0.75\textwidth]{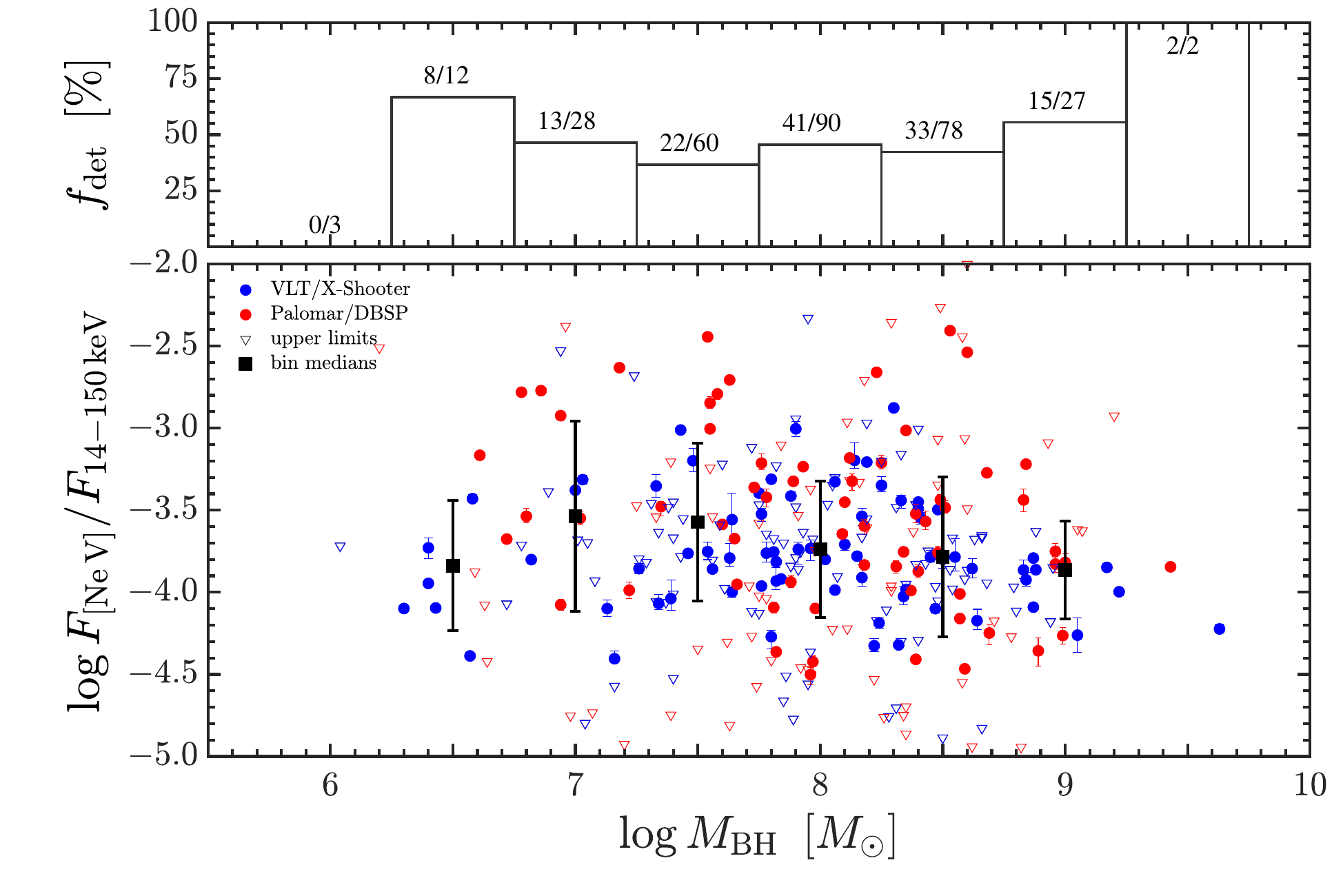}\\
    \includegraphics[clip, trim=1.5cm 0.0cm 0.0cm 0.1cm, width=0.75\textwidth]{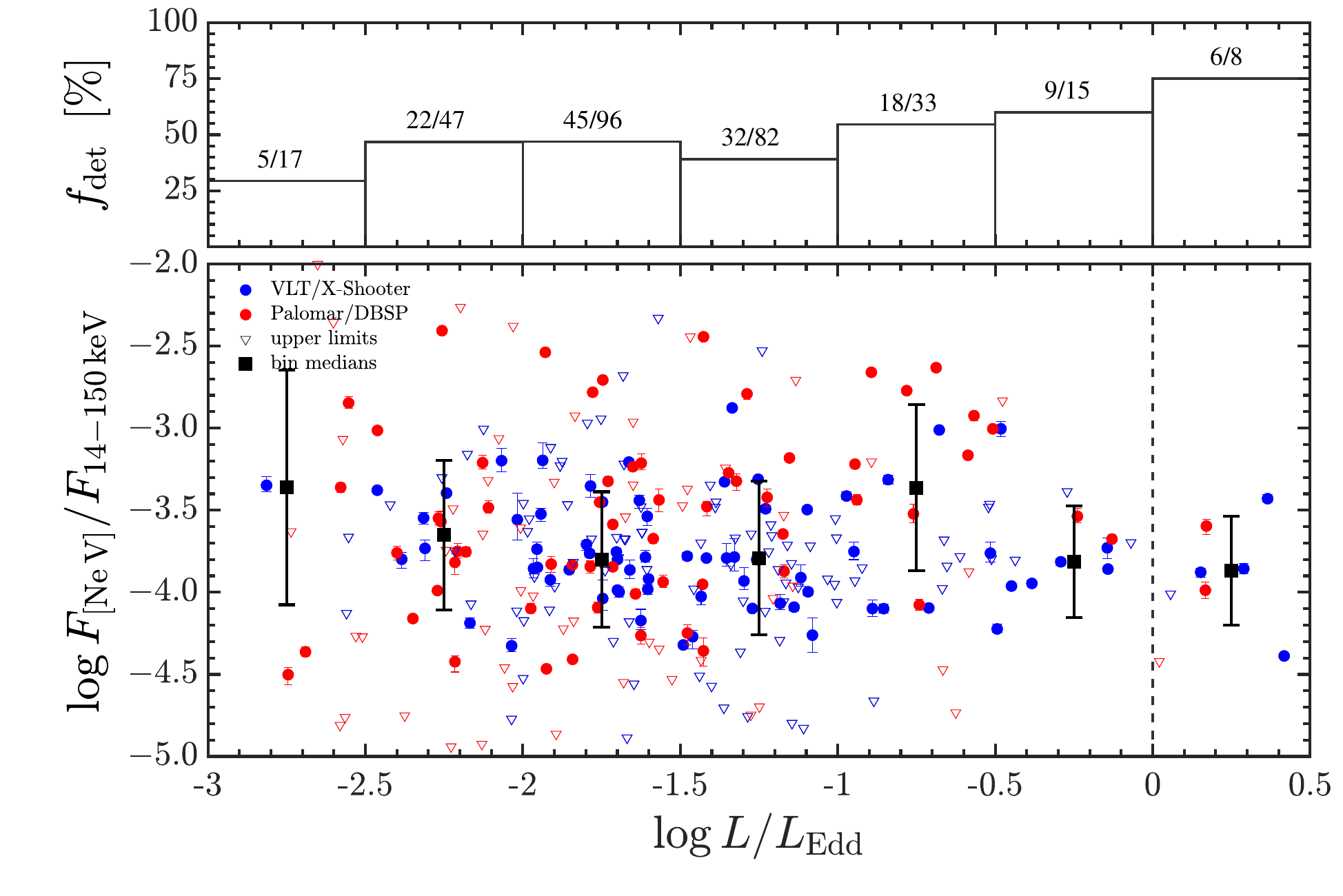}
    \caption{Same as Figure~\ref{fig:nev_vs_props_1}, but exploring trends between relative \nev\ line strength, \ffx, and key properties of our BASS sample deduced from their optical spectra, including black hole mass (\mbh, top) and Eddington ratio (\lamEdd, bottom). Here, too, we see a significant scatter and no evidence for strong trends in either \ffx\ or \fdet. Note in particular the lack of any evident trends towards the Eddington limit (vertical dashed line in bottom panel), although the scatter in \ffx\ is somewhat smaller.}
\label{fig:nev_vs_props_2}
\end{figure*}

We next look into how the relative strength of the \NeV\ emission line, i.e. \ffx, relates to several key AGN properties that are available for our sample through the BASS datasets.
Figures \ref{fig:nev_vs_props_1} and \ref{fig:nev_vs_props_2} show \ffx\ versus \Luhard, \NH, \mbh, and \lamEdd, while the smaller adjacent panels show how \fdet\ varies across each of these properties. As already noted, not all sources with measurements of, or upper limits on, \fnev\ (and thus \ffx) have the ancillary information needed to be considered for all the panels in Figs.~\ref{fig:nev_vs_props_1} \& \ref{fig:nev_vs_props_2}. Moreover, the main scatter plots focus on the range $-5\leq\log(\ffx)\leq-2$, which leaves out (at most) 24 sources with upper limits on \fnev\ (and thus on \ffx). 

Clearly, \ffx\ shows a considerable scatter across, and no evidence for  significant trends with, the newly considered properties shown in Figs.~\ref{fig:nev_vs_props_1} \& \ref{fig:nev_vs_props_2} (i.e., \NH, \mbh, and \lamEdd). The real scatter is yet higher than what is perceived in the main plots, given that some upper limits are found beyond the plots' limits.\footnote{Specifically, there are upper limits in the $\log(\ffx)<-5$ regime, spread across the range of properties shown.}
The lack of statistically significant correlations is confirmed through the appropriate correlation tests, which all yield $P>0.01$. The exact $P$-values are listed in Table~\ref{tab:corr_tests} in Appendix~\ref{app:corr_tests}. The only statistically significant relation among the properties presented in these plots is an anti-correlation between \ffx\ and \Luhard\ ($P_{\rm K}<0.01$), which is not surprising given the aforementioned sub-linear relation between \Lnev\ and \Luhard\ (Eq.~\ref{eq:Lnev_Luhard}).

We again note that alternative choices of bolometric corrections for estimating \lamEdd\ would not have changed our main findings, i.e. the large scatter in \ffx\ across the whole range of \lamEdd, and lack of statistically significant trends between these quantities, or between \fdet\ and \lamEdd. Moreover, given the large scatter we see in \ffx\ at any \mbh\ and \lamEdd, and the wide range our sample covers in \mbh\ and \lamEdd\ (roughly 3.5\,dex), it is highly unlikely that the (significant) systematic uncertainties related to these quantities ($\gtrsim$0.5\,dex) can hide a strong underlying (intrinsic) correlation between \ffx\ and either \mbh\ and/or \lamEdd.

Our findings regarding the links between \nev\ emission and key SMBH/AGN properties (or lack thereof) are in excellent agreement with the recent BASS study of the MIR \nev\ lines, by \cite{Bierschenk2024}.  In addition to the high detection rate of these lines across all levels of obscuration and the strong correlation between \nev\ and ultra-hard X-ray luminosities, that study found considerable scatter, and no trends between relative \nev\ line emission (i.e., \ffx) and neither \NH, \mbh, nor \lamEdd. In particular, there were no noticeable trends in \ffx\ towards the highest levels of obscuration (i.e., $\log[N_{\rm H}/\cmii]\gtrsim23.5$) and/or the highest Eddington ratios ($\lamEdd\approx1$).

On the other hand, the lack of observed links between \nev\ emission and neither \mbh\ nor \lamEdd\ for the BASS AGN (both here and in \citealt{Bierschenk2024}) stands in contrast with basic expectations from models of accretion flows onto SMBHs. Specifically, the radiation emitted from simple thin accretion disks is expected to become softer when \mbh\ increases, with disks around $\mbh\gtrsim10^9\,\Msol$ SMBHs expected to emit negligible amounts of radiation in the spectral regime relevant for \nev. Indeed, ``cold'' accretion disks were invoked to explain luminous quasars with weak (high ionization) emission lines \citep{LaorDavis2011}. In contrast, we find no trend of decreasing \ffx\ or \fdet\ with increasing \mbh, and \fdet\ remains significant even at $\mbh\gtrsim10^9\,\Msol$.
As mentioned in Section~\ref{sec:intro}, several models predict that super-Eddington accretion flow onto SMBH would have either enhanced \cite[e.g.,][]{Done2012,KubotaDone2019} or suppressed \cite[e.g.,][]{Pognan2020} extreme UV radiation, which at face value is expected to strongly affect \nev\ emission. In contrast, we find no evidence for a strong trend of either increasing nor decreasing \ffx\ (or \fdet) with \lamEdd. The scatter in \ffx\ does seem to decrease for $\lamEdd\gtrsim1$. In principle, this may be expected if the radiation from super-Eddington flows saturates at $\Lbol \approx {\rm few}\times L_{\rm Edd}$, as is suggested by many studies \cite[e.g.,][]{Abramowicz88,McKinney14}.
However, we caution that the statistical power of the $\lamEdd\gtrsim1$ AGN in our sample is too limited to draw any robust conclusions about this regime of accretion.

We next discuss in more detail the relations between \ffx\ or \fdet\ and \NH\ (or lack thereof; bottom panels of Fig.~\ref{fig:nev_vs_props_1}).
As noted above, we are able to detect \nev\ even in the highest-\NH\ systems, $\lognh\gtrsim23.5$, where \fdet\ seems to {\it increase}, exceeding $\fdet\approx50\%$ and reaching $\fdet\gtrsim80\%$ towards $\lognh>24$.
We stress, however, that there is no statistically significant correlation between \ffx\ and \NH. This latter finding is, again, in agreement with some previous studies of the MIR \nev\ lines \cite[e.g.,][]{Spinoglio2022,Bierschenk2024}, however it is in apparent contrast with what one may have concluded from other studies that investigated the (relative) strength of \nev\ in high-\NH\ sources.
Specifically, the study of \cite{Gilli2010} investigated links between \NeV\ and both X-ray emission and \NH, for several samples of AGN selected at various redshifts and through various methods. This includes a sample of 74 local AGN; 21 (highly) obscured AGN at $z\sim0.5$ from \citet[][drawn from a parent sample of \citealt{Zakamska2003}]{Vignali2010}; and a larger sample of unobscured, blue SDSS quasars at $z\sim0.1-1.5$, drawn from the catalog of \cite{Young2009}. The study by \cite{Gilli2010} highlights the prospect of using the \nev\ emission line, and particularly the \nev-to-X-ray ratio as a method to identify highly obscured (Compton-thick) AGN, specifically at intermediate redshifts, where \nev\ is accessible in the visual regime. Moreover, Figure~3 of \cite{Gilli2010} may be interpreted as hinting at a (weak) correlation of increasing \ewnev\ with increasing \NH, although we stress that the \cite{Gilli2010} study did not interpret it as such.

\begin{figure}
\centering
\includegraphics[clip, trim=1.5cm 0.0cm 0.0cm 0cm, width=0.475\textwidth]{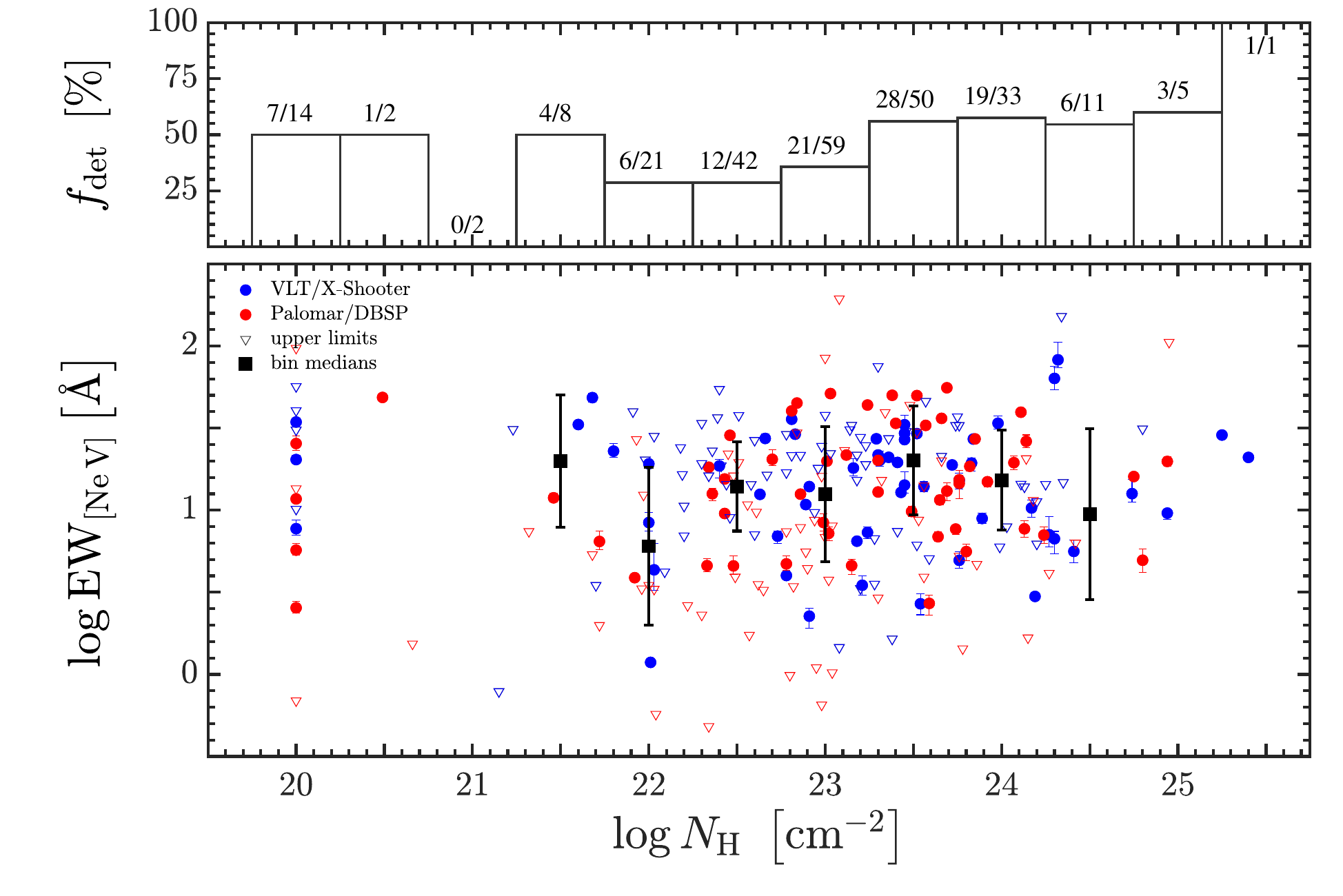}
    \caption{The relation between \ewnev\ and \NH. 
    Symbols are the same as in Figs.~\ref{fig:nev_vs_props_1} \& \ref{fig:nev_vs_props_2}.
    Here, too, the scatter is large and there is no evidence for a significant (positive) correlation between the quantities, in contrast to what may have been implied from similar plots in previous findings (see text for discussion).} 
\label{fig:ew_vs_nh}
\end{figure}

In this context, our BASS-based work confirms the ability to identify \nev\ emission in a significant fraction of highly obscured AGN, as noted above. 
However, when looking at the relation between \ewnev\ and \NH, which we show in Figure~\ref{fig:ew_vs_nh}, we again find a large scatter and lack of any obvious \& strong correlation between these quantities. We stress that our BASS sample is much larger, more homogeneously selected, and much more complete than the sample(s) used by \cite{Gilli2010}, where the (weak) trend between \ewnev\ and \NH\ was implied mostly through the combination of several very distinct samples, each covering markedly different \NH\ regimes (see above). We further point out that, for obscured AGN, \ewnev\ includes the host-dominated, rest-frame blue continuum. Both emission components (i.e. both the nominator and denominator of \ewnev) are subject to host-scale dust attenuation, but the stellar emission might be attenuated by dusty ISM that is out of the line-of-sight probed by the X-ray \NH\ measurements. Therefore, understanding any (potential) links between X-ray-deduced \NH\ and \ewnev\ requires knowledge of the morphology and gas content of the AGN hosts, which is complicated and expected to vary between individual sources, and across AGN samples. While our sample indeed focuses on obscured (i.e., narrow line) AGN, some of the samples in the \cite{Gilli2010} study are of unobscured, broad-line AGN. Such sources are naturally expected to have a low \NH, but also a low \ewnev\ at fixed \fnev\ (or \Lnev), as their continua are dominated by the quasar-like emission. Thus, combining samples of obscured and unobscured AGN may lead to an apparent correlation between \ewnev\ and \NH, however interpreting any such apparent correlations as evidence for {\it intrinsic} links between \nev\ emission and AGN obscuration remains challenging.

What our Fig.~\ref{fig:ew_vs_nh} does show is, again---and in agreement with \cite{Gilli2010}, that detecting \nev\ does not have to be more challenging for the most obscured AGN, and in fact can practically be even (slightly) easier, in terms of the contrast of the narrow \nev\ emission line with respect to the local continuum (which is what \ewnev\ measures).

\subsection{Stacking Analysis}
\label{sec:stack_res}

\begin{figure*}
\centering
\begin{minipage}[t]{0.48\textwidth}
\centering
\textbf{stacks of VLT/X-Shooter spectra} 
\includegraphics[width=1\textwidth]{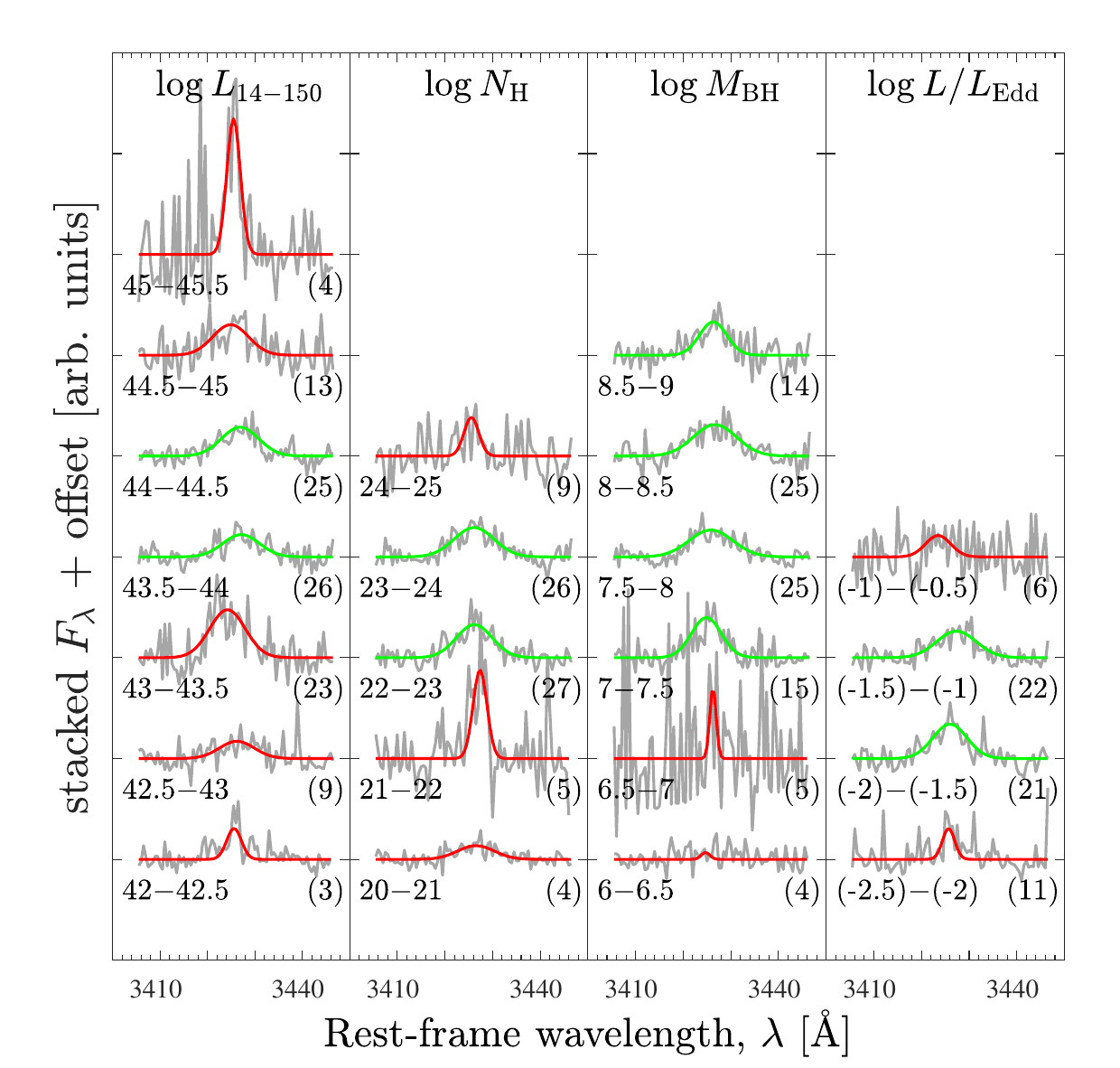}
\end{minipage}
\hfill
\begin{minipage}[t]{0.48\textwidth}
\centering
\textbf{stacks of Palomar/DBSP spectra} 
\includegraphics[width=1\textwidth]{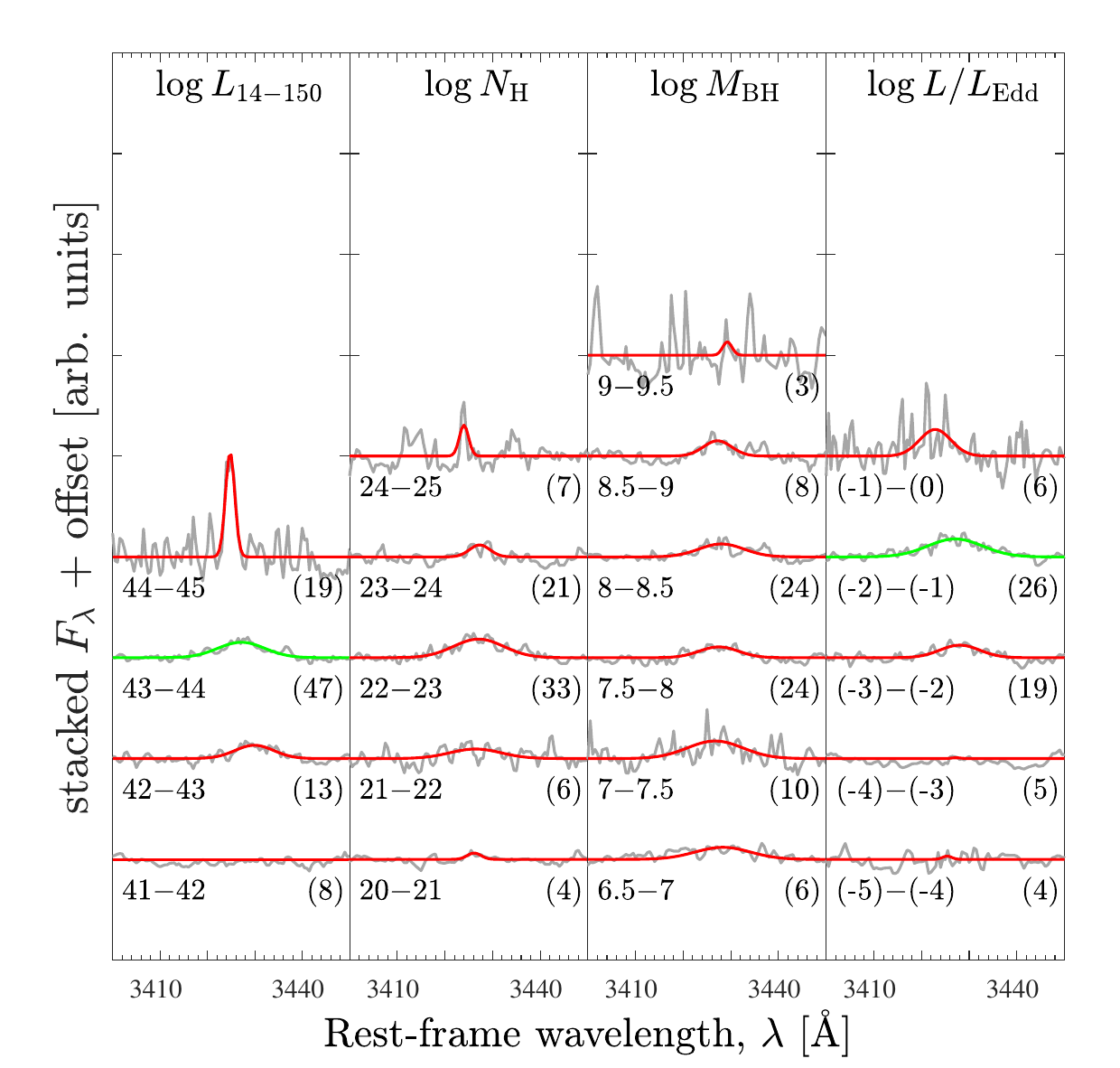}
\end{minipage}
    \caption{Results of spectral stacking analysis. Each column of panels traces spectral stacks in bins of a specific AGN property, as indicated (\Lx, \mbh, \NH, and \lamEdd). 
    The left- and right-hand-side sets of plots show stacks of X-Shooter and DBSP spectra, respectively.
    For each spectral stack, the annotation describes the range of the corresponding property and the number of spectra that belong to that bin (in square parentheses). The blue lines shows the corresponding stacked spectrum, while the colored lines trace the best-fitting (Gaussian) model of the stacked spectrum. Green lines denoting cases where the \nev\ line is robustly measured (continuum $S/N>3$, line $\ewnev/\Delta\ewnev > 3$ and $\fwhm<1200\,\kms$), while red lines denote the remaining cases. 
    As an example, the third panel from the top in the left-most column of the left set shows the stack of the 25 AGN with $44<\log(\Lx/\ergs)\leq44.5$ which were observed with X-Shooter. This specific stack resulted in a robust measurement of the \nev\ line.}
\label{fig:stacks}
\end{figure*}

Figure~\ref{fig:stacks} shows the stacked spectra we construct in various bins of \Luhard, \mbh, \NH, and \lamEdd, separately for spectra obtained with the VLT/X-Shooter (left panels) and Palomar/DBSP (right panels). For each bin of each of the properties and datasets we consider, we indicate the number of spectra that were used to construct the corresponding stack. We use line colors to indicate which stacks result in a robust narrow \nev\ line detection (see caption). These stacked spectra, and the analysis that follows, focus only on those AGN where \nev\ was not individually detected. This choice is motivated by our desire to look further into potential links between \nev\ (relative) strength and various AGN properties, focusing on those sources which were only considered as upper limits so far (i.e., in Figs~\ref{fig:nev_vs_props_1} \& \ref{fig:nev_vs_props_2}).
Table~\ref{tab:stacks} in Appendix~\ref{app:stack_tab} lists the basic \nev\ line measurements, particularly equivalent widths and their errors, or upper limits.

We stress again that here, too, we consider the {\it narrow} \nev\ line as detected in a stacked spectrum only if it can be robustly fit with a narrow ($\fwhm<1200\,\kms$) profile, over a robustly detected continuum level ($S/N>3$), and yielding a non-negligible ($\ewnev>0.01\,\AA$) and statistically significantly line emission (here, $\ewnev/\Delta\ewnev >3$).
As mentioned in Section~\ref{sec:spec_stack}, a visual inspection of the stacked spectra in Fig.~\ref{fig:stacks} suggests these criteria can be considered conservative, as there are several cases where the \nev\ can be visually identified, however not considered as robustly detected by our criteria. The clearer examples are those where the continuum is extremely noisier ($S/N\leq3$), and/or the line profile is broader ($\fwhm>1200\,\kms$) than what our criteria allow.

Figure~\ref{fig:stacks} and Table~\ref{tab:stacks} offer  several insights, despite the limitations mentioned above. First, the higher quality of the X-Shooter data (in terms of $S/N$) results in many more of the related stacks yielding robust \nev\ detections, even with a modest number of stacked spectra ($N_{\rm spec}\ge14$). By contrast, only two of the DBSP stacks result in a \nev\ detection---in some of the largest bins we have ($43<\log[\Luhard/\ergs)]\leq44$ with $N_{\rm AGN}=47$ sources; and $-2 < \log\lamEdd\leq-2$ with $N_{\rm AGN}=26$ sources).
Second, there is no apparent, clear trend between any of the (binned) AGN properties and the corresponding \nev\ stack, in terms of both the sheer detection of line emission, or its (relative) strength. Specifically, we can detect \nev\ in the intermediate-luminosity bins of X-Shooter spectra (the two bins covering $43.5<\log[\Luhard/\ergs]\leq44.5$), but not at higher or lower luminosities. 
Likewise, arrow \nev\ is robustly detected in the intermediate bins of both \lognh\ and \lamEdd, but not in the lowest or highest regimes of these quantities. While \nev\ is detected in the X-Shooter stacks of sources with $\log(\mbh/\Msun)>7$ (but not below this value), the \nev\ emission does not appear to become stronger with increasing \mbh.
The DBSP-based stacks obviously cannot contribute much to these insights.

We conclude that our stacking analysis generally supports the realization that neither the detection of \nev\ nor its relative strength are closely correlated with any of the key AGN quantities we've examined (\Luhard, \mbh, \lamEdd, and \NH). 
Since our stacking analysis is essentially based on \ewnev, the lack of trends between \ewnev\ and \NH\ in the stacked spectra further strengths our conclusion that the (statistically significant) correlation found between these quantities for individual spectra is not robust, and that the scatter between sources with comparable \NH\ is too large to consider this relation any further. Similarly, the (weak) anti-correlation found between \ffx\ and \Luhard\ (see Fig.~\ref{fig:nev_vs_props_1}) is not reflected in our stacking analysis, again likely due to the large scatter present in (relative) intrinsic \nev\ strength at fixed \Luhard, and the fact that the stacking analysis introduces further scatter by focusing on \ewnev\ (i.e., incorporates \nev\ strength relative to the host continuum, not AGN continuum).

%%%%%%%%%%%%%%%%%%%%%%%%%%%%%%%%%%%%%%%%%%%%%%%%%%%%%%%%%%%%%%%%

\subsection{What suppresses \nev\ emission in some AGN?}
\label{sec:dis_low_nev}

We have demonstrated that the narrow \NeV\ line remains (relatively) weak, or indeed absent, even among some of our higher-luminosity AGN, and/or in many of our high $S/N$ spectra. Moreover, the line remains undetected in many of our stacked spectra (Fig.~\ref{fig:stacks}). How can this be explained, given the lack of significant relations between neither \fdet\ nor \ffx\ and key AGN properties (Figs.~\ref{fig:nev_vs_props_1} and \ref{fig:nev_vs_props_2})?

Given the short wavelength of the \nev\ line, one explanation may have to do with the effects of dust on host galaxy scales. Indeed, there is some evidence that a significant fraction of BAT AGN host galaxies are gas-rich disks \cite[particularly the more massive ones; e.g.,][]{Koss2011_hosts,Koss2021_CO}, while other studies argued that host-scale gas may indeed significantly attenuate circumnuclear emission components \cite[see, e.g.,][and references therein]{BuchnerBauer2017,Gilli2022}.
Moreover, attenuation by dust is also expected to be higher in galaxy mergers, particularly during the late stages of major mergers \cite[e.g.,][]{Blecha2018}. Swift/BAT AGN provide strong evidence in support of such a link between mergers and dust attenuation \cite[][]{Ricci2017_mergers,Koss2018_mergers}, as well as the suppression of high-ionization line emission in mergers \cite[see, e.g.,][for the \oiii-to-X-ray ratio]{Koss2010_mergers}.
While a detailed analysis of the morphologies, gas content, and the attenuation of circumnuclear line emission regions in our sample is beyond the scope of the present study, we note again that even the highest-\NH\ sources among our AGN present a significant detection fraction of \nev\ emission (bottom panel of Fig.~\ref{fig:nev_vs_props_1}). We also note that if the 1\,dex (2\,dex) range we see in relative \nev\ intensity (i.e., range in \ffx\ at fixed) is to be explained by dust, it would require significant amounts of dust: either a  dust screen with $E(B-V)\approx0.4$ ($\approx$0.8), or $E(B-V)\approx0.8$ ($\approx$1.6) if the dust is mixed with emission region.
Furthermore, a quick visual inspection of the images of some of the systems that have exceptionally high $S/N$ but no \nev\ detection\footnote{Multi-band images drawn from the SDSS or Legacy Imaging Surveys web pages.} revealed no clear evidence for neither edge-on disk morphologies nor galaxy mergers.

Another way to explain the weakness, or indeed absence of \nev\ emission, may have to do with the physical properties of the line-emitting region. The recent study by \cite{McKaig2024} used radiative transfer calculations to examine how the (relative) strength of so-called ``coronal'' line emission depends on basic properties of the line-emitting gas and of the incident ionizing continuum. The presence of dust in the line-emitting region was shown to significantly suppress coronal line emission, at a given gas metallicity, driven by the depletion of metals onto dust grains. Although the dependence of \NeV\ strength on the metal and dust content of the gas was much milder than what was found for all other coronal lines studied in \cite{McKaig2024}, it was shown that it may be suppressed by as much as an order of magnitude, if the gas is assumed to be super-solar but dusty (compared with the dust-free scenario). Metallicity variations alone may naturally also change the (relative) line strength. The much stronger suppression of \NeV\ emission seen in some of the \cite{McKaig2024} models is driven by increasingly high gas densities which are implicit in their models of compact line-emitting regions (i.e., $\lesssim$0.1 pc for their fiducial AGN source, with $\Lbol = 10^{45}\,\ergs$). Since our study focuses on narrow \nev\ line emission, which originates further away from the central source, this feature of the \cite{McKaig2024} models is likely not relevant for interpreting our results.
Examining the physical properties of the gas-emitting regions of the BASS AGN and comparison to radiative transfer models may yield further insights regarding the mechanisms driving \nev\ emission in some AGN but not in others. However, this is far beyond the scope of the present work.

%%%%%%%%%%%%%%%%%%%%%%%%%%%%%%%%%%%%%%%%%%%%%%%%%%%%%%%%%%%%%%%%
\section{Summary and Concluding Remarks}
\label{sec:summary}

In this study, we have analyzed the narrow \NeV\ emission line in a large sample of low-redshift, ultra-hard X-ray selected AGN drawn from the BASS project. This allowed us to look for relations between \nev\ emission and other important AGN and SMBH properties, including AGN luminosities, SMBH masses and accretion rates (i.e., Eddington ratios), and line-of-sight gas column densities (i.e., circumnuclear obscuration)---both across our sample of individual measurements and in stacked spectra.
Our main conclusions are as follows:

\begin{itemize}

    \item 
    The narrow \NeV\ emission line was robustly detected in roughly half of the AGN for which we were able to look for it ($43\%$, \Ndet/\Nnev\ sources). Given our somewhat conservative criteria for defining robust \nev\ detections, this can be considered a lower limit on the fraction of narrow-line BASS AGN that exhibit this emission line. \NeV\ can be detected in a significant fraction of even the most obscured AGN in our sample, with detection rates exceeding $\fdet\sim60\%$ for column densities reaching $\lognh \simeq24$. On the other hand, even some of our highest-$S/N$ spectra yield no robust \nev\ detection, with $\fdet\lesssim70\%$ even for spectra with $S/N\gtrsim50$. 
    See Figures~\ref{fig:detper}, \ref{fig:nev_vs_props_1} and \ref{fig:nev_vs_props_2}, and Section~\ref{sec:detect_frac}.

    \item 
    Among the \nev-detected sources, the typical scaling between the \nev\ line emission originating from the narrow and/or coronal line region(s), and the ultra-hard X-ray emission related to the central engine, is $\Lnev/\Luhard/\simeq1/5000$ (median denominator: 5700, mean: 4600, standard deviation: 0.45\,dex). For the more commonly used X-ray emission in the $2-10\,\kev$ range we find $\Lnev/\Lhard\simeq1/2200$ (median denominator: 2274, mean: 2044, standard deviation: 0.47\,dex; again, among the \nev\ detections). We caution that the underlying, intrinsic scatter on these scaling factors is larger, given the numerous sources for which \nev\ was not detected.
    See Figures~\ref{fig:nev_vs_bat} \& \ref{fig:nev_xray_ratio}, and Section~\ref{sec:LL_scaling}.

    \item 
    Combining these scalings with the simplistic, commonly-used universal bolometric correction of $\Lbol/\Luhard=8$ one can in principle obtain estimates of \Lbol\ based on \NeV\ measurements, namely $\Lbol\approx45,000\times\Lnev$ (i.e., a logarithmic scaling factor of 4.65\,dex; based on the median $\Lnev/\Luhard$ ratio).
    Assuming $\Lbol/\Lhard=20$ and using the median $\Lnev/\Lhard$ ratio instead, would lead to an essentially identical (approximate) scaling, however the real scatter here is somewhat larger. 

    \item 
    We find no strong evidence for meaningful links between neither the \nev\ detection rate (\fdet) nor its relative strength (\ffx) and any of the key AGN properties we examined, including AGN luminosity (\Luhard), BH mass (\mbh), Eddington ratio (\lamEdd), and column density (\NH). While there is a statistically significant anti-correlation between \ffx\ and \Luhard, the trend is weak and the scatter is large even when considering only the \nev-detected sources. Given the large number of non-detections, we caution against over-interpreting any weak trend that one may see in     Figures~\ref{fig:nev_vs_props_1}, \ref{fig:nev_vs_props_2}, and \ref{fig:ew_vs_nh}. 
    See Section~\ref{sec:props_trends}.

    \item 
    Our stacking analysis provides further evidence against any significant and/or strong trends between (relative) \nev\ strength and the key AGN properties we examined.
    See Figure~\ref{fig:stacks} and Section~\ref{sec:stack_res}.
    
\end{itemize}

The \NeV\ emission line was long considered a robust tracer of AGN activity, given the exceptionally high ionization level required to produce it. Our results indeed strengthen this notion, and are in good agreement with several previous studies, including those that focused on the MIR lines of \nev\ \cite[e.g.,][]{Weaver2010,Spinoglio2022,Bierschenk2024}. Our work further emphasizes the usability of \NeV\ in identifying (highly) obscured AGN, and studying their accretion power, again in line with earlier studies \cite[e.g.,][]{Gilli2010,Vignali2010, Vergani2018,Li2024_NeV}. 

There are several directions of study enabled by the ability to detect the \NeV\ line and link it to intrinsic AGN accretion power (i.e., the \ffx\ scaling we quantified). Perhaps most importantly, the \nev\ line can be used to robustly identify (obscured) AGN in the extremely early Universe, thanks to the unprecedented capabilities of JWST. Specifically, the NIRSpec instrument \citep{Jakobsen22_NIRSpec,Boker23_NIRSpec} can probe the spectral region of the \nev\ emission line for $z\simeq0.8-14.5$ sources. For example, if a $z=9$ source would exhibit a \nev\ line with $F\approx3\times10^{-19}\,\ergcms$, which is comparable with the deepest line detections in the JWST Advanced Deep Extragalactic Survey \citep[JADES;][]{Eisenstein2023_JADES_overview}, the corresponding line luminosity of $\Lnev\approx3\times10^{41}\,\ergs$, combined with our fiducial scaling relation ($\log[\ffx]=-3.75$) and a bolometric correction of $\Lbol = 8\times \Luhard$ would imply $\Lbol\approx1.4\times10^{46}\,\ergs$. While this is comparable with many known luminous $z\simeq6$ quasars \cite[e.g.,][and references therein]{Fan2023_review}, the potential to detect \emph{obscured} AGN at $z>7$ is indeed novel.
A forthcoming publication (by Trakhtenbrot et al.) will explore the usability of the scaling relations established in the present study for some of the deepest JWST data available for high redshift sources.

Second, high spatial resolution imaging with narrow bands finely chosen to cover \NeV\ (e.g., with HST/STIS) might be used to reveal dual AGN with sub-kpc separations in (highly) obscured and/or disturbed galaxy nuclei, where such dual AGN are expected to be particularly common \cite[e.g.,][]{Koss2018_mergers,Koss2023_dual}. Unlike other, perhaps more accessible lines like \OIII, which may be powered by complex circumnuclear SF structures, the \NeV\ line could serve as an unambiguous tracer of (dual) AGN photoionization. A limitation here would be the high levels of obscuration, which are expected in (late-stage) mergers \citep[e.g.,][]{Ricci2017_mergers,Blecha2018}, similarly to what has been found for \oiii\ diagnostics \cite{Koss2010_mergers}. Finally, as noted in Section~\ref{sec:spec_line} and demonstrated in Fig.~\ref{fig:outflows}, some of our AGN exhibit outflow signatures in their \nev\ line profiles. The \nev\ line may thus serve as a unique tracer of AGN-related outflows, in lieu of the more common practice of studying such phenomena by much more challenging analysis involving line ratio diagnostics for the outflow-related parts of the line profiles.

The intrinsic weakness of the \NeV\ line resulted in over half of our sample to lack individual detections, which somewhat limited the statistical power of our study. It also naturally limits the usability of \nev\ for studies of obscured, high redshift and/or dual AGN. The high redshift case might be particularly affected by the potentially low metal abundance in the earliest galaxies, as well as the possibility that extreme stellar populations, or other non-AGN mechanisms, would also produce significant \nev\ emission. 
Recent progress in the theoretical understanding of high-ionization line emission in both AGN and non-AGN sources \cite[e.g.,][]{Simmonds2021,Cleri2023_diag,McKaig2024} may help better understand, and ultimately alleviate, some of these issues. 

Line emission from highly ionized species, such as the \NeV\ line, combined with benchmark studies of large, well understood AGN samples, such as BASS, can therefore allow to directly probe some of the key stages in SMBH and galaxy (co-)evolution, including the kind of AGN and SMBH populations that are still missing from our census of accreting SMBHs.

%%%%%%%%%%%%%%%%%%%%%%%%%%%%%%%%%%%%%%%%%%%%%%%%%%%%%%%%%%%%%%%%
\acknowledgments

We thank the anonymous referee for their invaluable comments, which helped us improve the paper in several important aspects.
We thank D.\ Baron for her help with partial correlation analyses, and J.\ McKaig for useful discussions.

T.R.\ and B.T.\ acknowledge support from the European Research Council (ERC) under the European Union's Horizon 2020 research and innovation program (grant agreement number 950533) and from the Israel Science Foundation (grant number 1849/19).
C.R.\ acknowledges support from Fondecyt Regular grant 1230345, ANID BASAL project FB210003 and the China-Chile joint research fund.
M.K.\ acknowledges support from NASA through ADAP award 80NSSC22K1126.
K.O.\ acknowledges support from the Korea Astronomy and Space Science Institute under the R\&D program (Project No. 2025-1-831-01), supervised by the Korea AeroSpace Administration, and the National Research Foundation of Korea (NRF) grant funded by the Korea government (MSIT; RS-2025-00553982).
Y.D. acknowledges financial support from a Fondecyt postdoctoral fellowship (3230310).

This research was supported by the Excellence Cluster ORIGINS which is funded by the Deutsche Forschungsgemeinschaft (DFG, German Research Foundation) under Germany's Excellence Strategy - EXC 2094 - 390783311.
B.T. acknowledges the hospitality of the Instituto de Estudios Astrof\'isicos at Universidad Diego Portales, and of the Instituto de Astrof\'isica at Pontificia Universidad Cat\'olica de Chile.

\bigskip
\facilities{VLT:XSHOOTER, Palomar:Hale, Swift(BAT)}

\software{MAAT \citep{Ofek2014_matlab}, astropy \citep{Astropy2013,Astropy2018}}

\clearpage
\newpage

\appendix

\section{Outflow signatures}
\label{app:outflows}

\begin{figure*}
\centering
\includegraphics[width=0.495\textwidth]{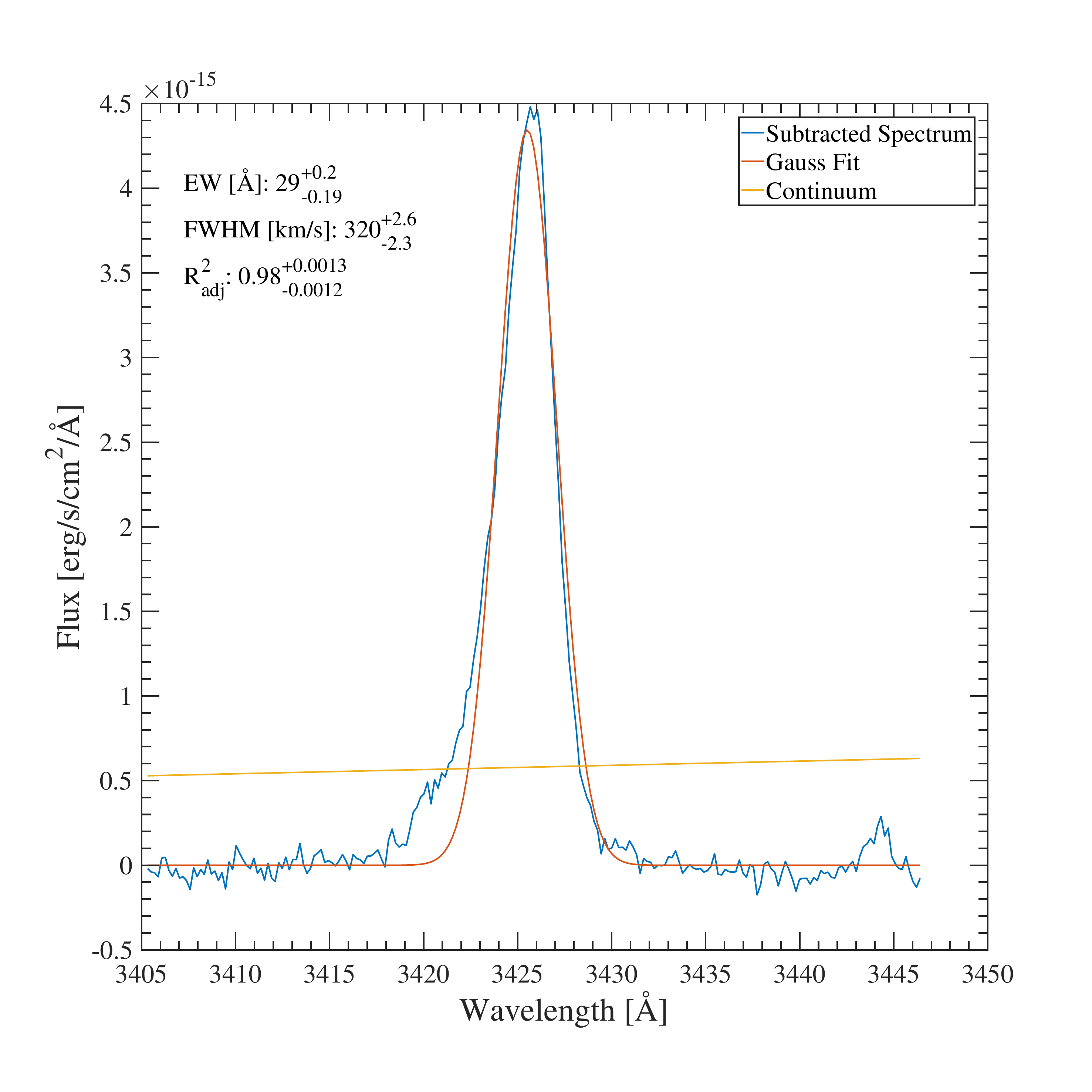}\hfill
\includegraphics[width=0.495\textwidth]{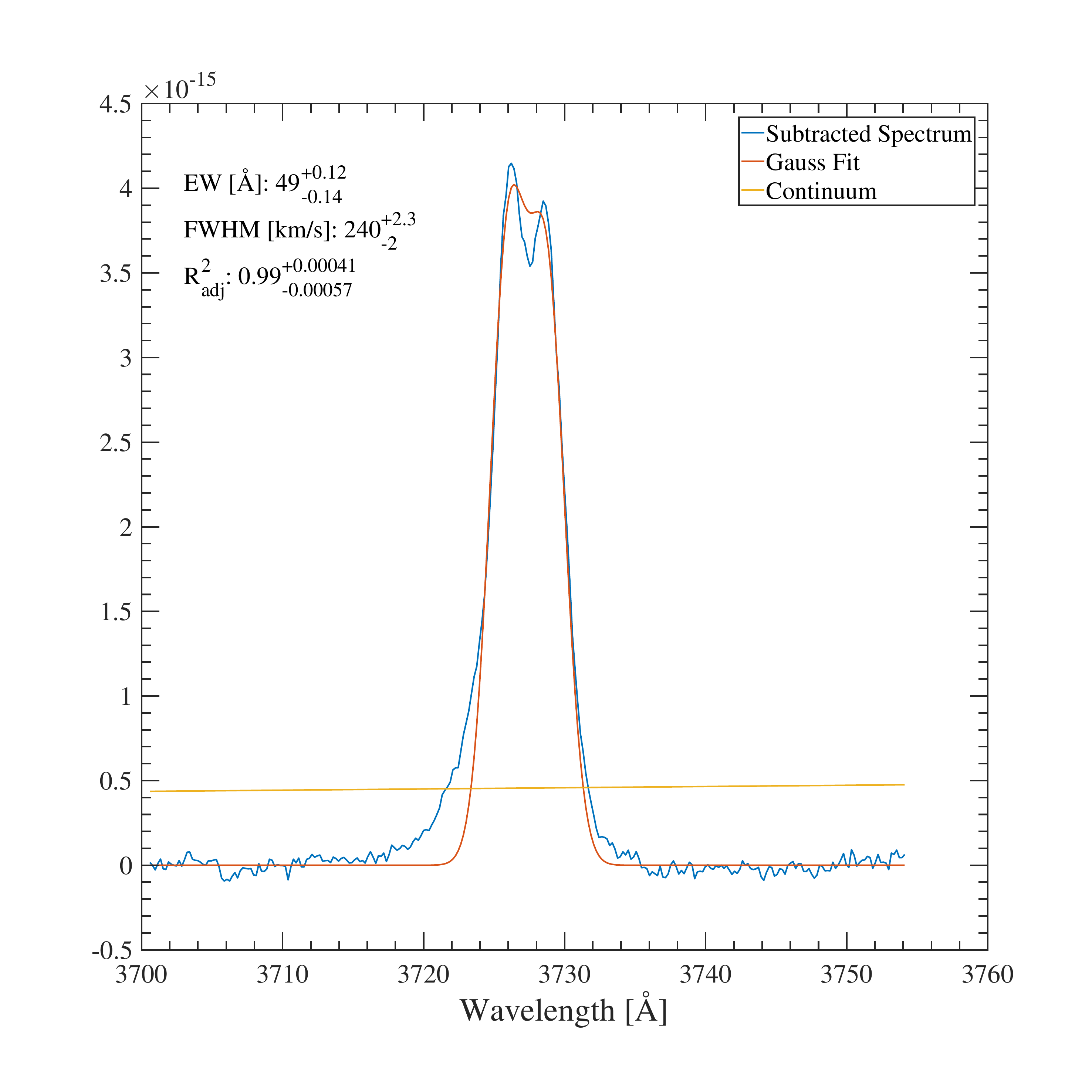}
\caption{Example of outflow signatures in the \NeV\ (left) and \OII\ (right) emission lines, for the AGN BASS ID 1138. In both panels, the yellow lines mark the original continuum level (subtracted from the observed spectrum), while the red lines represent our best-fitting line profiles (we used two Gaussians to model the \oii\ doublet). At face value, the shape of the prominent \oii\ profile in an AGN would indicate that this AGN-hosting galaxy has an outflow, however this outflow could have been photoionized by young stars, and have little to do with the AGN itself. The fact that the \nev\ line (also) shows an outflow signature immediately indicates an AGN-photoionized outflow, which makes the system all the more relevant for discussing co-evolutionary scenarios.} 
\label{fig:outflows}
\end{figure*}

In Figure~\ref{fig:outflows} we show an example of an AGN with markedly asymmetric profiles of both its \NeV\ and \OII\ emission lines. The \OII\ fitting was performed through a procedure that is essentially identical to the one used for \NeV. The fact that a blue wing is seen in \nev\ and not only in \oii\ indicates that the (implied) outflowing gas is photoionized by the AGN, and not by any host-scale SF processes.

\section{Spectral Measurements from Stacked Spectra}
\label{app:stack_tab}

Table~\ref{tab:stacks} presents the \nev\ spectral measurements for the stacked spectra, in various bins of \Luhard, \mbh, \lamEdd, and \NH, separately for stacks based on X-Shooter and DBSP data. 
For simplicity, we tabulate the $\ewnev$ and $\Delta\ewnev$ measured for every stacked spectrum, however several other quantities are required to determine whether a narrow \nev\ line is robustly detected (see Section~\ref{sec:spec_stack} for details).

%%%%%%%%%%%%%%%%%%%%%%%%%%%%%%%%%%%%%%%%%%%%%%%%%%%%%%%%%%%%%%%%

\begin{deluxetable*}{lc|ccc|ccc}
\label{tab:stacks}
\tablecaption{NeV stacking analysis results}
\tablewidth{\textwidth}
\tablehead{
\colhead{Property} & \colhead{Bin range} &
\multicolumn{3}{c}{VLT/X-Shooter} & 
\multicolumn{3}{c}{Palomar/DBSP} \\ 
\colhead{} & \colhead{} & 
\colhead{$N_{\rm AGN}$} & \colhead{det.?\tablenotemark{$\ast$}} & \colhead{EW [\AA]} &  
\colhead{$N_{\rm AGN}$} & \colhead{det.?\tablenotemark{$\ast$}} & \colhead{EW [\AA]} 
}
\startdata
%%%%%%%%%%%%%%%%%%%%%%%%%%%%%%%%%%%%%%%%%%%%%%%%%%%%%%%%%%%%%%%%%%%%%%%%%%%%%%%%%%%%
%%%%%%%%%%%%%%%%%%%%%%%%%%%%%%%%%%%%%%%%%%%%%%%%%%%%%%%%%%%%%%%%%%%%%%%%%%%%%%%%%%%%
\hline
$\log(\Luhard/\ergs)$ &   $(40-41]$ &  0 &      & $-$       &                  2  &      & $\ll 0.1$             \\
~~~~~~~~~~~~~~~~~~~~~~~~~~ &   $(41-42]$ &  1 &      & $-$       &                  8  &      & $\ll 0.1$             \\
%%%%%%%%%%%%%%%%%%%%%%%%%%%%%%%%%%%%%%%%%%
~~~~~~~~~~~~~~~~~~~~~~~~~~ & $(42-42.5]$ &  3 &      & $2.3\pm1.8$ & \multirow{2}{*}{13} &      & \multirow{2}{*}{$2.6\pm1.2$} \\
~~~~~~~~~~~~~~~~~~~~~~~~~~ & $(42.5-43]$ &  9 &      & $2.5\pm1.7$ &                     &      &                              \\
%%%%%%%%%%%%%%%%%%%%%%%%%%%%%%%%%%%%%%%%%%
~~~~~~~~~~~~~~~~~~~~~~~~~~ & $(43-43.5]$ & 23 &      & $5.8\pm2.1$ & \multirow{2}{*}{47} & \multirow{2}{*}{\chm} & \multirow{2}{*}{$3.7\pm0.9$} \\
~~~~~~~~~~~~~~~~~~~~~~~~~~ & $(43.5-44]$ & 26 & \chm & $4.4\pm1.0$ &                     &                       &                  \\
%%%%%%%%%%%%%%%%%%%%%%%%%%%%%%%%%%%%%%%%%%
~~~~~~~~~~~~~~~~~~~~~~~~~~ & $(44-44.5]$ & 25 & \chm & $4.6\pm0.9$ & \multirow{2}{*}{19} &      & \multirow{2}{*}{$3.9\pm2.0$}      \\
~~~~~~~~~~~~~~~~~~~~~~~~~~ & $(44.5-45]$ & 13 &      & $3.0\pm2.2$ &                     &      &                                   \\
%%%%%%%%%%%%%%%%%%%%%%%%%%%%%%%%%%%%%%%%%%
~~~~~~~~~~~~~~~~~~~~~~~~~~ & $(45-45.5]$ &  4 &      & $4.6\pm1.8$ & \multirow{2}{*}{1}  &      & \multirow{2}{*}{$-$}              \\
~~~~~~~~~~~~~~~~~~~~~~~~~~ & $(45.5-46]$ &  0 &      & $-$         &                     &      &                                   \\
~~~~~~~~~~~~~~~~~~~~~~~~~~ &  $(46-47]$  &  0 &      & $-$         &                 1   &      & $-$                               \\
%%%%%%%%%%%%%%%%%%%%%%%%%%%%%%%%%%%%%%%%%%%%%%%%%%%%%%%%%%%%%%%%%%%%%%%%%%%%%%%%%%%%
%%%%%%%%%%%%%%%%%%%%%%%%%%%%%%%%%%%%%%%%%%%%%%%%%%%%%%%%%%%%%%%%%%%%%%%%%%%%%%%%%%%%
\hline
$\log(\NH/\cmii)$  & $(20-21]$ &  4	&      & $2.6\pm1.7$	&	4	&	&	$0.5\pm0.3$	\\
~~~~~~~~~~~~~~~~~  & $(21-22]$ &  5	&	   & $4.7\pm3.3$	&	6	&	&	$3.1\pm2.0$ \\
~~~~~~~~~~~~~~~~~  & $(22-23]$ & 27	& \chm & $5.0\pm0.9$	&	33	&	&	$3.8\pm1.4$ \\
~~~~~~~~~~~~~~~~~  & $(23-24]$ & 26	& \chm & $5.0\pm1.0$	&	21	&	&	$1.6\pm1.0$	\\
~~~~~~~~~~~~~~~~~  & $(24-25]$ &  9	&	   & $2.9\pm2.1$	&	7	&	&	$1.5\pm1.1$	\\
%%%%%%%%%%%%%%%%%%%%%%%%%%%%%%%%%%%%%%%%%%%%%%%%%%%%%%%%%%%%%%%%%%%%%%%%%%%%%%%%%%%%
%%%%%%%%%%%%%%%%%%%%%%%%%%%%%%%%%%%%%%%%%%%%%%%%%%%%%%%%%%%%%%%%%%%%%%%%%%%%%%%%%%%%
\hline
$\log(\mbh/\Msol)$ & $(6-6.5]$ &  4	&		 &	$0.9 \pm1.0$ &	2	&	&	$5.2\pm0.2$ \\
~~~~~~~~~~~~~~~~~  & $(6.5-7]$ &  5	&		 &	$1.1 \pm0.9$ &	6	&	&	$3.7\pm2.3$ \\
~~~~~~~~~~~~~~~~~  & $(7-7.5]$ & 15	&	\chm &	$5.2 \pm1.2$ &	10	&	&	$2.8\pm2.4$ \\
~~~~~~~~~~~~~~~~~  & $(7.5-8]$ & 25	&	\chm &	$5.3 \pm1.1$ &	24	&	&	$2.0\pm1.3$ \\
~~~~~~~~~~~~~~~~~  & $(8-8.5]$ & 25	&	\chm &	$6.0 \pm1.2$ &	24	&	&	$2.2\pm1.6$ \\
~~~~~~~~~~~~~~~~~  & $(8.5-9]$ & 14	&	\chm &	$4.7 \pm1.2$ &	8	&	&	$2.2\pm2.0$ \\
~~~~~~~~~~~~~~~~~  & $(9-9.5]$ &  1	&		 &	$-$	         &	3	&	&	$0.7\pm1.4$ \\
%%%%%%%%%%%%%%%%%%%%%%%%%%%%%%%%%%%%%%%%%%%%%%%%%%%%%%%%%%%%%%%%%%%%%%%%%%%%%%%%%%%%
%%%%%%%%%%%%%%%%%%%%%%%%%%%%%%%%%%%%%%%%%%%%%%%%%%%%%%%%%%%%%%%%%%%%%%%%%%%%%%%%%%%%
\hline
$\log\lamEdd$      & $((-5.0)-(-4.0)]$  &  0 &     & $-$         &   4                 & &	$0.1\pm0.1$ \\
~~~~~~~~~~~~~~~~~  & $((-4.0)-(-3.0)]$  &  0 &     & $-$         &   5                 & &   $\ll 0.1$    \\
%%%%%%%%%%%%%%%%%%%%%%%%%%%%%%%%%%%%%%%%%%
~~~~~~~~~~~~~~~~~  & $((-3.0)-(-2.5)]$ &  1	&      & $-$         & \multirow{2}{*}{19} & & \multirow{2}{*}{$2.6\pm1.5$} \\
~~~~~~~~~~~~~~~~~  & $((-2.5)-(-2.0)]$ & 11	&	   & $1.8\pm1.2$ &                     & &                              \\
%%%%%%%%%%%%%%%%%%%%%%%%%%%%%%%%%%%%%%%%%%
~~~~~~~~~~~~~~~~~  & $((-2.0)-(-1.5)]$ & 21	& \chm & $5.6\pm1.6$ & \multirow{2}{*}{26} & \multirow{2}{*}{\chm} & \multirow{2}{*}{$4.9\pm1.1$}  \\
~~~~~~~~~~~~~~~~~  & $((-1.5)-(-1.0)]$ & 22	& \chm & $5.0\pm1.0$ &                     &                       &             \\
% %%%%%%%%%%%%%%%%%%%%%%%%%%%%%%%%%%%%%%%%%%
~~~~~~~~~~~~~~~~~  & $((-1.0)-(-0.5)]$ &  6	&   & $1.3\pm1.6$ & \multirow{2}{*}{6} & & \multirow{2}{*}{$1.9\pm2.6$}  \\
~~~~~~~~~~~~~~~~~  & $((-0.5)-( 0.0)]$ &  2	&	& $-$	      &                    & &                         \\
~~~~~~~~~~~~~~~~~  & $((0.0)-(+0.5)]$  &  1	&	& $-$         & \multirow{2}{*}{2} & & \multirow{2}{*}{$-$}    \\
~~~~~~~~~~~~~~~~~  & $((+0.5)-(1.0)]$  &  0	&	& $-$         &                    & &                         \\
%%%%%%%%%%%%%%%%%%%%%%%%%%%%%%%%%%%%%%%%%%%%%%%%%%%%%%%%%%%%%%%%%%%%%%%%%%%%%%%%%%%
%%%%%%%%%%%%%%%%%%%%%%%%%%%%%%%%%%%%%%%%%%%%%%%%%%%%%%%%%%%%%%%%%%%%%%%%%%%%%%%%%%%%
% \hline
\enddata
\tablenotetext{\ast}{The symbol \chm\ marks the stacked spectra where significant narrow \nev\ line emission was detected.}
\end{deluxetable*}

\clearpage
%%%%%%%%%%%%%%%%%%%%%%%%%%%%%%%%%%%%%%%%%%

\clearpage

\onecolumngrid

\section{Results of correlation tests}
\label{app:corr_tests}

Table~\ref{tab:corr_tests} lists the results of the hypothesis tests we conducted in search for correlations between (relative) \nev\ emission strength and various other properties.

\begin{deluxetable*}{llcccccc}
\label{tab:corr_tests}
\tablecaption{Results of Correlation Tests}
\tablewidth{\textwidth}
\tablehead{
 \multicolumn2c{Variables} & 
 \multicolumn2c{Spearman} & 
 \multicolumn2c{Pearson} & 
 \multicolumn2c{Kendall} \\ 
 \colhead{$Y$} & \colhead{$X$} & \colhead{$\rho_{S}$} & \colhead{$P_{S}$} & \colhead{$\rho_{P}$} & \colhead{$P_{P}$} & \colhead{$\tau_{K}$} & \colhead{$P_{K}$}
} 
% \decimals
\startdata
 $\bm{\log \fnev}$ & $\bm{\log \Fuhard}$ & $+0.44$ & $\bm{<10^{-5}}$ & $+0.47$ & $\bm{<10^{-6}}$ & $+0.19$ & $\bm{<10^{-5}}$ \\
 $\bm{\log \Lnev}$ & $\bm{\log \Luhard}$  & $+0.80$ & $\bm{<10^{-6}}$ & $+0.83$ & $\bm{<10^{-6}}$ & $+0.38$ & $\bm{<10^{-6}}$ \\
 $\bm{\log \ffx}$ & $\bm{\log \Luhard}$  & $-0.34$ & $\bm{3.1\times10^{-5}}$ & $-0.35$ & $\bm{1.4\times10^{-5}}$ & $-0.11$ & $\bm{0.008}$ \\ 
 $\log \ffx $ & $\log\mbh$      & $-0.17$ & 0.04 & $-0.15$ & 0.08 & $-0.06$ & 0.17 \\ 
 $\log \ffx $ & $\log\lamEdd$   & $-0.15$ & 0.11 & $-0.16$ & 0.10 & $-0.05$ & 0.28 \\ 
 $\log \ffx $ & $\log\NH$       & $-0.02$ & 0.51 & $-0.05$ & 0.46 & $+0.06$ & 0.22 \\
 $\log \ewnev$ & $\log\NH$      & $+0.06$ & 0.51 & $+0.08$ & 0.40 & $+0.05$ & 0.47 \\
\enddata
\tablecomments{Statistically significant correlations  (i.e., where all three correlation tests resulted in $P<0.01$) are highlighted in boldface.}
\end{deluxetable*}

%%%%%%%%%%%%%%%%%%%%%%%%%%%%%%%%%%%%%%%%%%

% %%%%%%%%%%%%%%%%%%%%%%%%%%%%%%%%%%%%%%%%%%%%%%%%%%%%%%%%%%%%%%%%%
\clearpage
\newpage
%%%%%%%%%%%%%%%%%%%%%%%%%%%%%%%%%%%%%%%%%%%%%%%%%%%%%%%%%%%%%%%%%

\bibliography{BASS_NeV}{}
\bibliographystyle{aasjournal}

\end{document}

% End of file `sample63.tex'.